%% file: binary-only.tex
\documentclass[prd, aps, twocolumn, nofootinbib, showpacs, superscriptaddress,longbibliography]{revtex4-1}
\usepackage[utf8]{inputenc}
\usepackage{graphicx,mathtools,xcolor,latexsym,rotating,soul}
\usepackage[T1]{fontenc}
\usepackage{tikz}
\usepackage{comment}
\usepackage{soul}
\usepackage{amsmath, amssymb, amsthm, amsfonts,breqn}
\usepackage{multirow}
\usetikzlibrary{decorations.pathmorphing}
\usepackage[colorlinks=true,citecolor=green,urlcolor=blue,linkcolor=blue]{hyperref}
\usepackage{pifont}
\usepackage{gensymb}
\usepackage{enumitem}

\newtheoremstyle{named}{}{}{\itshape}{}{\bfseries}{.}{.5em}{\thmnote{#3 }#1}
\theoremstyle{named}

\newcommand*{\rom}[1]{\expandafter\@slowromancap\romannumeral #1@}

\newcommand{\stf}[1]{\langle #1 \rangle}

\newcommand{\AB}{A \leftrightarrow B}
\newcommand{\FI}{f \leftrightarrow i}

\makeatletter
\let\cat@comma@active\@empty
\makeatother
\begin{document}
\title{Probing internal dissipative processes of neutron stars \\ with gravitational waves during the inspiral of neutron star binaries}

\author{Justin L. Ripley}
\email{ripley@illinois.edu}
\affiliation{Illinois Center for Advanced Studies of the Universe, Department of Physics, University of Illinois at Urbana-Champaign, Urbana, IL 61801, USA}

\author{Abhishek Hegade K. R.}
\email{ah30@illinois.edu}
\affiliation{Illinois Center for Advanced Studies of the Universe, Department of Physics, University of Illinois at Urbana-Champaign, Urbana, IL 61801, USA}

\author{Nicol\'as Yunes}
\email{nyunes@illinois.edu}
\affiliation{Illinois Center for Advanced Studies of the Universe, Department of Physics, University of Illinois at Urbana-Champaign, Urbana, IL 61801, USA}

\begin{abstract}
We study the impact of out-of-equilibrium, dissipative effects on the dynamics of inspiraling neutron stars.
We find that modeling dissipative processes (such as those from the stars internal effective fluid viscosity) requires that one introduce a new tidal deformability 
parameter--the dissipative tidal deformability--which modifies the phase of gravitational waves emitted during the inspiral phase of a neutron star binary. 
We show that the dissipative tidal deformability corrects the gravitational-wave
phase at 4 post-Newtonian order for quasi-circular binaries.
This correction receives a large finite-size enhancement by the stellar compactness, analogous to the case of the tidal deformability.
Moreover, the correction is not degenerate with the time of coalescence, which also enters at 4PN order, 
because it contains a logarithmic frequency-dependent contribution.
Using a simple Fisher analysis, we show that physically allowed values for the dissipative tidal deformability may be constrained by measurements of the phase of emitted gravitational waves to 
roughly the same extent as the (electric-type, quadrupolar) tidal deformability.
Finally, we show that there are no out-of-equilibrium, dissipative corrections to the tidal deformability itself.
We conclude that there are at least two relevant tidal deformability parameters that 
can be constrained with gravitational-wave phase measurements during the 
late inspiral of a neutron star binary:
one which characterizes the adiabatic tidal response of the star, and another which
characterizes the leading-order out-of-equilibrium, dissipative tidal response. 
These findings open a window to probe dissipative processes in the interior of neutron stars with gravitational waves.
\end{abstract}
\maketitle
\section{Introduction\label{sec:introduction}}

A long-standing goal in astrophysics and nuclear particle physics has been 
to determine the neutron star equation of state \cite{Ozel:2016oaf,Lattimer:2021emm,Burgio:2021vgk}, i.e., the relation between the pressure and the energy density in the interior of a neutron star.
The neutron star equation of state affects a variety of macroscopic and, in principle, observable properties of neutron stars, such as their mass, radius, moment of inertia, and tidal deformability. The latter is a measure of how much a star deforms in response to an exterior tidal field induced, for example, by its companion in a binary system (for a review see~\cite{Poisson-Will}). The tidal deformations of stars in a binary affect their orbital evolution, and therefore, these deformations become encoded in the emitted gravitational waves~\cite{Flanagan:2007ix,Read:2009yp,Hinderer:2009ca}.   
The measurement of the tidal deformability of neutron stars is challenging, as tidal corrections enter the
waveform at 5 post-Newtonian (PN) order\footnote{A term in the mathematical representation of 
some quantity is said to be of $N$ PN order if it scales as $(v/c)^{2N}$ relative to the 
leading-order piece of that representation in a PN expansion; for a review see \cite{Poisson-Will}.}
\cite{Flanagan:2007ix}. 
Despite this, measurements by the LIGO-Virgo-KAGRA  collaboration have already begun to
place the first meaningful constraints on the tidal deformability, and thus, on the equation of state of neutron stars 
with gravitational waves~\cite{LIGOScientific:2018cki,LIGOScientific:2018hze,
Annala:2017llu,Bauswein:2017vtn,Most:2018hfd,Raithel:2018ncd,De:2018uhw,Chatziioannou:2018vzf,Carson:2018xri,
LIGOScientific:2019eut,LIGOScientific:2020aai}.

The tidal deformability of a star is usually defined in terms of its tidal Love number, 
which describes the linear and \textit{adiabatic} (or ``prompt'') response of the star with respect to an imposed external 
gravitational field \cite{love-original-article,Hinderer:2009ca,Damour:2009vw,Binnington:2009bb}. Recently, extensions beyond this quasi-static response have begun to be investigated. 
One such extension is the study of ``dynamical tides,'' meant to represent f-mode-like harmonic oscillations of a star, 
which may be excited very close to merger~\cite{Hinderer:2016eia,Steinhoff:2016rfi,Schmidt:2019wrl}\footnote{Even if the 
f-mode is not resonantly excited during the inspiral, it may still bias the measurement of the tidal Love
number \cite{Pratten:2021pro}.}.  
Here we consider a different extension of the quasi-static tidal response of a star to an external perturbation: 
the non-adiabatic, 
out-of-equilibrium, and \textit{dissipative} (or just dissipative for short) tidal
response of a neutron star. 
After reviewing an effective-field-theory approach to define this quantity, we derive its impact on the phase of gravitational waves emitted 
during late inspiral but before the merger of a neutron star binary. 

Before reviewing earlier work, we first outline how dissipative effects impact the dynamics of two initially non-spinning stars in a binary. 
The gravitational potential of one star induces a tidal bulge in the other star, and the presence of viscosity makes the bulge slightly lag the slowly changing gravitational potential, as the two objects orbit each other. 
This lag then torques the star, which causes it to start spinning. 
This spinning motion requires energy and angular momentum, which is extracted from the orbital angular momentum of the system, thus accelerating the rate of inspiral.

It is well known that these effects play an important role in the tidal dynamics of giant stars and satellites (for reviews see \cite{Zahn:2008fk,Ogilvie_2014}).
For giant stars, the effect of tidal dissipation is so large that it ultimately locks the spins of the objects in a binary system to the orbital frequency of the binary.
One of the first attempts at quantifying the impact of dissipative effects on the gravitational-wave inspiral phase of neutron star binaries was performed by Bildsten and Cutler in \cite{1992ApJ...400..175B}.
In that work, the authors estimated how viscosity would affect the binary through the process of \textit{dissipative tidal spin up}. 
Bildsten and Cutler found that the dissipative (through viscous forces) timescale for momentum transport across a neutron star would need to be of the order of the light-crossing time of the star in order for this effect to tidally lock the stars, and thus appreciably affect the binary evolution\footnote{Earlier work on the tidal interactions of neutron star binaries found similar results \cite{1992ApJ...398..234K}.}.
Such a short viscous transport timescale would require values of shear/bulk viscosity that are unphysically large, which led them to conclude that viscous tidal spin up cannot be detected with gravitational-wave observations. 
This conclusion (see also \cite{1992ApJ...398..234K}) led to the general expectation that viscous effects should be unmeasurable during the inspiral\footnote{We note that non-viscous/non-dissipative effects can also spin-up a neutron star as it orbits it companion, although these effect are also expected to be too small to be measurable \cite{Lai:1993di}.}.

In this work, we revisit the calculation of tidal dissipation in neutron star binaries.
In our analysis, we consider a general parametrization of the tidal response of a neutron star that includes its leading-order dissipative response. 
We ultimately find that, even in the absence of tidal locking, dissipative effects enter the gravitational-wave phase of quasi-circular binaries at 4PN order. 
By itself, this result is not new; tidal dissipative effects are known to appear at 4PN order in black hole binaries \cite{Poisson:1994yf,Tagoshi:1997jy,Alvi:2001mx}. 
Unlike earlier work though, we find that values of tidal dissipation much smaller than those needed to tidally lock a neutron star before merger could lead to an observable imprint on the gravitational wave phase of neutron star binaries. 
This essentially arises because the tidal dissipation parameter that enters the phase receives a large finite-size enhancement (analagous to the case for the tidal Love number).  

We introduce a new ``dissipative'' tidal deformability parameter that characterizes the leading-order viscous correction to the gravitational-wave phase.
This dissipative tidal deformability parameter has yet to be  self-consistently calculated for relativistic stars. 
Nevertheless, we provide an order-of-magnitude estimate of its value, which leads us to conclude that physically allowed values of dissipative (viscous) effects could potentially be constrained with current gravitational-wave ground-based detectors.
By ``physically allowed'' values of dissipation, we mean values of dissipation that are small enough that the average timescale of dissipative momentum transport across the star is less than the speed of light \cite{1992ApJ...400..175B}. 
We note that many different physical processes could in principle contribute to the dissipative tidal deformability. 
The largest contributions contributions for neutron stars are expected to arise from Urca processes, which can lead to
an effective bulk viscosity for sufficiently cold star
~\cite{Alford:2017rxf,Most:2021zvc,Most:2022yhe}.
We perform a preliminary Fisher analysis to estimate the ability to constrain such mechanisms using advanced LIGO with a GW170817 like event.
We leave a more rigorous Bayesian analysis to future work.

Finally, we show that there are no dissipative corrections to the tidal Love number itself.
Thus, the measurement of two tidal deformability parameters--one which enters at
4PN and one which enters at 5PN in the gravitational-wave phase--characterizes the dominant dissipative and equilibrium properties of matter at supra-nuclear densities inside neutron stars.

The remainder of the paper presents the detailed derivation that leads to the conclusions summarized above, and is organized as follows. 
We introduce an effective point particle action in Sec.~\ref{sec:effective-point-particle}, where we show that out-of-equilibrium effects enter the relativistic equations of motion at 6.5 PN order.
We then specialize to the Newtonian dynamics of two objects in Sec.~\ref{sec:viscosity-PN-order-phase}, where we derive the Newtonian equations of motion, and then derive the
quadrupolar gravitational radiation emitted from the binary within the adiabatic, quadrupolar approximation.
In Sec.~\ref{sec:regime-of-applicability}
we turn to the self-consistency of the approximations we made in deriving the gravitational-wave phase in 
Sec.~\ref{sec:viscosity-PN-order-phase} and present results from a Fisher analysis.
In Sec.~\ref{sec:no-out-of-equlibrium-contribution-tidal-love}, we demonstrate that there
are no out-of-equilibrium contributions to the tidal Love number.
We conclude and point to future research in Sec.~\ref{sec:conclusions}.
The appendices review our notation, and include more details of the fluid models we use
in Sec.~\ref{sec:no-out-of-equlibrium-contribution-tidal-love}.
Our notation for fluid variables follows~\cite{Rezzolla-Book}.

\section{Effective point particle description of compact objects and notation\label{sec:effective-point-particle}}
In this section, we model the tidal interaction of a compact binary system immersed in an external field using a 
point-particle action \cite{1989thyg.book..128D}.
In the effective field theory description of extended objects,
finite size effects are accounted for via higher-derivative
terms in the action \cite{Damour:1995kt,Damour:1998jk,Goldberger:2004jt} (for a more recent review see \cite{Porto:2016pyg}).
To leading order in derivatives, the effective point-particle action reads 
\cite{Goldberger:2005cd}
\begin{align}
    \label{eq:effective_action} 
    S
    =
    \int d\tau \left(
        -
        mc^2
        +
        Q_{(E)}^{\mu\nu}E_{\mu\nu}
        +
        Q_{(B)}^{\mu\nu}B_{\mu\nu}
        +
        \cdots
    \right)
    ,
\end{align}
where $\tau$ is the proper time of the particle, 
$E_{\mu\nu}\equiv C_{\mu\alpha\nu\beta}v^{\alpha}v^{\beta}$ and 
$B_{\mu\nu}\equiv\frac{1}{2}\epsilon_{\mu\alpha\beta\rho}C^{\alpha\beta}{}_{\nu\sigma}v^{\rho}v^{\sigma}$ are the electric and magnetic parts of the
Weyl tensor respectively, $Q_{(E/B)}^{\mu\nu}$ are the object's electric/magnetic quadrupole
moments, and $v^{\mu}\equiv dx^{\mu}/d\tau$ is its four-velocity. 

The response of the body to the external field can be complicated and non-linear.
For weak enough tidal fields, a simple, albeit phenomenological, way to model the response is through a linear response function. For concreteness, we focus on the electric part of the Weyl tensor and write 
\begin{align}
    \label{eq:linear_response_q_f_e}
    Q^{\mu\nu}_{(E)}\left(\tau\right)
    =
    \int^{\infty}_{-\infty} 
        d\tau^{\prime} 
        \left(F_2\right)^{\mu\nu}_{\alpha\beta}\left(\tau-\tau^{\prime}\right)
        E^{\alpha\beta}\left(\tau^{\prime}\right)
    .
\end{align}
From here on we consider a homogeneous, retarded transfer function
\begin{align}
    \left(F_2\right)^{\mu\nu}_{\alpha\beta}
    \left(\tau-\tau^{\prime}\right)
    =
    \delta^{\mu}_{\alpha}\delta^{\nu}_{\beta}
    \Theta\left(\tau-\tau^{\prime}\right)
    F_2\left(\tau-\tau^{\prime}\right)
    ,
\end{align}
where $\Theta\left(\tau-\tau^{\prime}\right)$ is the step function.
Matching the internal dynamics of the compact object to its external motion amounts to determining the response function $F_2\left(\tau-\tau^{\prime}\right)$~\cite{Chakrabarti:2013xza}.

We can simplify our parameterization of the linear response function by using a separation of scales argument.
If the orbital timescale is much longer than the internal, 
dynamical timescales of the star, we make a common assumption and assert that the tidal response can be written as
\begin{align}
    \label{eq:general_time_expansion}
    Q^{\mu\nu}_{(E)}\left(\tau\right)
    =&
    -
    \lambda^{(0)}_2
    \sum_{n=0}^{\infty} \tau^{(n)}_2 
    \frac{d^n}{d\tau^n}E^{\mu\nu}\left(\tau\right)
    .
\end{align}
This expansion implies that $F_2\left(\tau-\tau^{\prime}\right)$ is analytic in Fourier space. 
That is, it is equivalent to $F_2\left(\omega\right)$ taking the form (c.f. \cite{Chakrabarti:2013xza}),
\begin{align}
    \label{eq:fourier-transform}
    F_2\left(\omega\right)
    =
    -
    \lambda_2\sum_{n=0}^{\infty}\tau_2^{(n)}\left(i\omega\right)^n
    .
\end{align}
We note that $\lambda_{2}^{(0)}$ has SI units of $\mathrm{meter}^5$ 
and $\tau_{2}^{(d)}$ has SI units of $\mathrm{seconds}^d$.

The coefficients $\tau_2^{(n)}$ quantify the response of the quadrupole moment to the changes in the tidal field. We can set $\tau_2^{(0)}=1$ through a definition of $\lambda_2^{(0)}$, and we will do so for the remainder of this article.
Truncating the expansion of the (electric) quadrupole moment to first order, we obtain
\begin{align}
    \label{eq:linear_time_expansion}
    Q^{\mu\nu}_{(E)}
   =&
    -
    \lambda_2^{(0)}E^{\mu\nu}
    -
    \lambda_2^{(0)} \tau_2^{(1)}\frac{d}{d\tau}E^{\mu\nu}
    \nonumber\\
    &
    + 
    \mathcal{O}\left( \tau_2^{(2)} \left|\frac{d^2E_{\mu \nu}}{d\tau^2}\right|\right)
    .
\end{align}
We see that $\lambda_2^{(0)}$ is proportional to the adiabatic tidal Love number
\cite{Hinderer:2007mb,Damour:2009vw,Binnington:2009bb}.
By contrast, the term proportional to $\tau_2^{(1)}$ 
captures dissipative tidal effects, 
and introduces a time delay between the value of $Q_{\mu\nu}$ and $E_{\mu\nu}$ \cite{1981A&A....99..126H}.

We now simplify notation by setting $\tau_2^{(1)} = -\tau_d$ and discuss why this quantity captures the tidal delay.
Viscous effects are known to cause a time delay between the
induced quadrupole moment of a body and the time-dependent, imposed gravitational field.
\cite{1879RSPT..170....1D,1981A&A....99..126H}
Assuming that the delay time $\tau_d$ is roughly constant, we then have
\begin{align}
    \label{eq:motivation-dissipation-negative}
    \mathcal{Q}_E^{\mu\nu}\left(\tau\right)
    =
    -
    \lambda_2^{(0)}E^{\mu\nu}\left(\tau-\tau_d\right)
    .
\end{align}
Series expanding this expression in $\tau_d$, we can identify $\tau_2^{(1)}=-\tau_d$.

Before continuing, we compare our effective action [Eq.~\eqref{eq:effective_action}]
with another model that is commonly used to model ``dynamical tides'', i.e.,~the impact of neutron star oscillations
on the orbital dynamics \cite{Steinhoff:2016rfi}. 
In this approach, the point particle is modeled as a simple harmonic oscillator that oscillates at the lowest (fundamental) frequency of the star.
Considering only the electric-type deformation of the neutron star, the effective
point particle Lagrangian in that case reads 
(for quasi-Newtonian treatments see also \cite{Flanagan:2007ix,Poisson-Will})
\begin{align}
    \label{eq:effective_action_sho}
    \mathcal{L}_{DT} 
    =&
    \frac{z}{4\lambda\omega_f^2}\left( 
        \frac{c^2}{z^2}
        \frac{dQ_{\mu\nu}^{(E)}}{d\tau}\frac{dQ^{\mu\nu}_{(E)}}{d\tau}
        -
        \omega_f^2Q_{\mu\nu}^{(E)}Q^{\mu\nu}_{(E)} 
    \right)
    \nonumber\\
    &-
    \frac{z}{2}E_{\mu\nu}Q_{(E)}^{\mu\nu}
    ,
\end{align}
where $z^2\equiv v_{\mu}v^{\mu}$ is the redshift factor,
$\omega_f$ is the lowest oscillatory frequency of the star 
(the $f$-mode), and $\lambda$ is a constant.
Varying the action and transforming to frequency space,
we can arrange the equation of motion for $Q^{\mu\nu}_{(E)}$ to read
\begin{align}
    \label{eq:Q_freq_sho}
    Q^{\mu\nu}_{(E)}
    =&
    -
    \frac{
        \lambda
    }{
        1 + \frac{c^2}{z^2}\frac{1}{\omega_f^2}\frac{d^2}{d\tau^2}
    }
    E^{\mu\nu}
    \nonumber\\
    =&
    -
    \lambda \sum_{n=0}^{\infty}
        \left(-1\right)^n
        \left(\frac{c^2}{z^2\omega_f^2}\frac{d^2}{d\tau^2}\right)^n
    E^{\mu\nu}
    .
\end{align}
We see that Eq.~\eqref{eq:Q_freq_sho} takes the form of
Eq.~\eqref{eq:general_time_expansion}, with $\tau_2^{(n)}=0$ if $n$ is odd.
We note that Eq.~\eqref{eq:effective_action_sho} does not take into account dissipative effects, 
because the action is clearly time-reversal invariant.

Finally, we provide an estimate of the regime of the validity of our model, 
represented through Eq.~\eqref{eq:linear_time_expansion},
for the evolution of an object in a bound gravitational binary.
We replace $d/d\tau \to 1/\tau_{sys}$, where $\tau_{sys}$ is a
characteristic proper timescale of the object's motion.
Later on, when carrying out a more quantitative estimate, 
we will set $\tau_{sys}$ to be the orbital period for binary motion.
Following our above discussion, 
we set $\tau_{2}^{(1)}$ to be equal to the characteristic
tidal lag timescale $\tau_d$, and $\tau_2^{(2)}$ to be equal to the square of the characteristic internal
oscillatory proper timescale of the object, $\tau_{int}^2\approx \omega_f^{-2}$. 
Inserting this into Eq.~\eqref{eq:general_time_expansion}, we obtain the inequalities
\begin{align}\label{eq:tower-time-inequalities}
    1 
    &\gg 
    \left|\frac{\tau_d}{\tau_{sys}}\right| \,,\quad
    1
    \gg 
    \left(\frac{\tau_{int}}{\tau_{sys}}\right)^2
    \,.
\end{align}
The expansion Eq.~\eqref{eq:general_time_expansion}
is valid provided these inequalities are satisfied\footnote{We 
note that as $\tau_d$ and $\tau_{int}$ capture distinct physical effects ($\tau_d$ describes
purely dissipative effects, while $\tau_{int}$ is conservative), we do not have
the inequality $|\tau_d/\tau_{sys}| \gg \left(\tau_{int}/\tau_{sys}\right)^2$.
In effect, if $1\gg\left(\frac{\tau_{int}}{\tau_{sys}}\right)^2$ holds, then $\lambda_2^{(0)}$
describes the leading-order conservative response of the system; we are not near a resonance.
If $1 \gg \left|\frac{\tau_d}{\tau_{sys}}\right|$ holds, then we only need to expand to linear
order in $d/d\tau$ to describe the dissipative effects in the tidal response.}.
For a quasi-circular binary orbit, this breaks down only right before merger, 
when the characteristic time lag timescale and the oscillatory timescale become comparable to each other and to the orbital period. 
We verify this explicitly for one phenomenological parameterization in Sec.~\ref{sec:inequality-verification}.
\section{Out-of-equilibrium effects enter the gravitational-wave phase at 4PN order \label{sec:viscosity-PN-order-phase}}
In this section, we derive the leading-order contribution to the gravitational-wave phase that depends on $\lambda_2^{(1)}$.
We set up our notation in Sec.~\ref{sec:notation-love}, and 
in Sec.~\ref{sec:Newt-eom}, we show that the equations of motion for a point
particle are modified at 6.5PN order by dissipative
corrections arising from $\tau_2^{(1)}$. 
Despite this, we show in Sec.~\ref{sec:gw-phase} that the gravitational-wave phase is affected
by dissipative effects at 4PN order.
In short, this is because dissipative effects decrease the Newtonian orbital energy of the binary, 
while the adiabatic tidal Love numbers conserve the Newtonian orbital energy.
\subsection{Notation}\label{sec:notation-love}
We label the two objects in the binary, and any physical
quantities associated with them, with the subscripts $A$ and $B$. We use the word ``object'' instead of ``neutron star'' or ``black hole'',
to remain agnostic about the system in question.
The masses of the two objects are $m_A$ and $m_B$, 
the total mass is $M\equiv m_A+m_B$, the reduced mass is 
$\mu\equiv m_Am_B/M$, and the symmetric mass ratio is $\eta\equiv \mu/M$.
The characteristic radii of the objects are $R_A$ and $R_B$, which could refer to the equatorial radius for a neutron star or the areal radius for a black hole. 
We also define the mass ratio by $q = m_A/m_B$.
The compactness of body $A$ is defined by 
\begin{align}
    C_A &\equiv \frac{G m_A}{R_A c^2}\,.
\end{align}
The response coefficient $\lambda_{2,A}^{(0)}$ 
of object $A$ is related to the tidal Love number $k_{2,A}$ by
\begin{align}\label{eq:lambda2A}
    \lambda_{2,A}^{(0)} 
    =
    \frac{2}{3} k_{2,A} R_A^5
    \,.
\end{align}
We define two dimensionless tidal deformabilities
\begin{subequations}
\label{eq:definition-tidal-deformabilities}
\begin{align}
    \Lambda_{A}
    &\equiv
    \frac{\lambda_{2,A}^{(0)} c^{10}}{(G m_A)^5}
    =
    \frac{2}{3}
    \frac{k_{2,A}}{C_{A}^5}
    ,\\
    \label{eq:Lambda12A}
    \Xi_{A}
    &\equiv
    -\frac{\lambda_{2,A}^{(0)} \tau^{(1)}_{2,A} c^{13}}{(G m_A)^6}
    =
    \frac{2}{3}
    \frac{k_{2,A}}{C_{A}^6}
    \frac{c\tau_{d,A}}{R_A}
    .
\end{align}
\end{subequations}
The parameter $\Lambda_{A}$ is the (electric-type, quadrupolar) 
tidal deformability, and is sometimes denoted by $\bar{\lambda}_{A}$ in the gravitational-wave literature.
We call $\Xi_A$ the dissipative tidal deformability.
We also define two binary tidal deformabilities 
\begin{subequations}\label{eq:lambdabar-xibar}
\begin{align}
    \label{eq:bar-Lambda}
    \bar{\Lambda} 
    &
    \equiv
    f(\eta)\frac{\Lambda_A + \Lambda_B}{2}
    +
    g(\eta)\frac{(\Lambda_A - \Lambda_B)}{2}
    \,, \\
    \label{eq:bar-Xi}
    \bar{\Xi} &
    \equiv
    f_1(\eta) \frac{\Xi_A + \Xi_B}{2}
    +
    g_1(\eta) \frac{\Xi_A - \Xi_B}{2} 
    \,,
\end{align}
\end{subequations}
where
\begin{subequations}
\begin{align}
    f(\eta) &= \frac{16}{13}
    \left(1 + 7 \eta - 31 \eta^2 \right) \,,\\
    g(\eta) &= -\frac{16}{13}
    \sqrt{1 - 4 \eta} 
    \left( 1 + 9 \eta - 11 \eta^2\right)\,,
    \\
    f_1(\eta)
    &= 
    8\left(2 \eta^2-4 \eta +1\right) 
    \,,
    \\
    g_1(\eta) 
    &=
    -8\sqrt{1-4 \eta } (1-2\eta)
    \,.
\end{align}
\end{subequations}
The binary tidal deformabilities $\bar{\Lambda}$~\cite{Hinderer_2010,Chatziioannou_2020} and $\bar{\Xi}$ 
will appear in the gravitational waveform instead of the individual tidal deformabilities of each body.
We work with Cartesian coordinates $\left(ct, x, y, z \right)$.
The indices $\left(i,j,\ldots\right)$ are used to denote spatial coordinates only.
We denote the Newtonian quadrupolar moment of body $A$ by $I^{ij}_{A}$.
Note that $I^{ij}_A$ is the non-relativistic limit of $Q^{ij}_A$.
We occasionally denote time derivatives with an overhead dot.
We use $\stf{\cdots}$ in index lists to denote the symmetric trace-free combination of tensorial indices.
\subsection{Newtonian equations of motion}\label{sec:Newt-eom}
We work in the center-of-mass frame, where the equations of motion 
for a binary take on a simple form (for a review, see \cite{Poisson-Will}). 
To leading order in the multi-polar moments of objects $A$
and $B$, the center-of-mass equations of motion are
\begin{align}
\label{eq:newtonian-eob-acceleration}
    a_i
    =
    &
    -
    \frac{GM}{r^2}n_i
    +
    \frac{GM}{2}\left(
        \frac{I_A^{\stf{jk}}}{m_A}
        +
        \frac{I_B^{\stf{jk}}}{m_B}
    \right)
    \partial_i\partial_j\partial_k\frac{1}{r}
    ,
\end{align}
where $a^i\equiv \ddot{x}^i_A - \ddot{x}^i_B$ is the relative acceleration.
Notice that we have \emph{not} included the spin of the objects in the equations
of motion; we discuss the validity of neglecting the spin induced on each object 
due to tidal torquing in Sec.~\ref{sec:regime-applicability}.

The Newtonian, linear tidal response of body $A$ is given by the non-relativistic limit of 
Eq.~\eqref{eq:linear_time_expansion}, namely
\begin{align}
    \label{eq:newtonian-first-order-limit-Iij}
    G \; I^{ij}_A
    \approx
    \left(\frac{Gm_A}{c^2}\right)^5
    \left(
        \Lambda_A\mathcal{E}_{A}^{ij}
        -
        \frac{Gm_A}{c^3}
        \Xi_A
       \dot{\mathcal{E}}_A^{ij}  
    \right)
    ,
\end{align}
where $\mathcal{E}_A^{ij}$ is the quadrupolar tidal field felt by $A$, as caused by $B$,
\begin{align}
    \label{eq:tidal-field-EA}
    \mathcal{E}_A^{ij}
    &=
    -
    Gm_B\partial^{i}\partial^{j}\frac{1}{r}
    =
    -
    \frac{3Gm_B}{r^3}n^{\stf{ij}}
    .
\end{align}
The time derivative of $\mathcal{E}_A^{ij}$ is
\begin{align}
    \dot{\mathcal{E}}_A^{ij}
    &=
    \frac{9Gm_B}{r^4}\dot{r}n^{\stf{ij}}
    -
    \frac{3Gm_B}{r^3}\left(\dot{n}^in^j+n^i\dot{n}^j\right)
    \nonumber\\
    &=
    \frac{9Gm_B}{r^4}\left[
        \dot{r}n^{\stf{ij}}
        -
        \frac{2}{3}\left(
            v^{(i}n^{j)}
            -
            \dot{r}n^in^j
        \right)
    \right]
    ,
\end{align}
where we have used the relative velocity $v^i\equiv \dot{x}^i_A - \dot{x}^i_B$,
and $ n_{\stf{ij}}\equiv n_in_j-\frac{1}{3}\delta_{ij}$.
Note that $\dot{\mathcal{E}}^{ij}_A$ is a symmetric trace-free tensor because $n_iv^i = \dot{r}$.
We conclude that the quadrupolar moment of body $A$ is
\begin{align}
\label{eq:quadrupole-moment-Iij}
    I^{ij}_A
    =
    &
    \frac{3m_B}{r^3}
    \left(\frac{Gm_A}{c^2}\right)^5
    \Bigg[
        \left(
            \Lambda_{A}
            +
            \frac{3Gm_A}{c^3}\frac{\dot{r}}{r}\Xi_{A}
        \right)
        n^{\stf{ij}}
        \nonumber\\
        &
        \qquad\qquad\qquad
        -
        \frac{2Gm_A}{c^3} 
        \frac{v^{(i}n^{j)} - \dot{r}n^in^j}{r}
        \Xi_{A}
    \Bigg]
    \,,
\end{align}
with analogous expressions for object $B$. 

With this in hand, we find that the center-of-mass equations of motion are
\begin{align}
    \label{eq:eob-equations-of-motion-expanded}
    a^j
    &
    =
    - 
    \frac{G M}{r^2} \Bigg\{
        n^j 
        \nonumber\\
        &
        +
        \frac{9 G^5 }{c^{10}r^5}
        \left[ 
            m_B m_A^4 \Lambda_{A} 
            + 
            \AB
        \right]
        n^j
        \nonumber \\
        &
        +
        \frac{9 G^6}{c^{13}r^6}
        \left[
            m_B m_A^5 \Xi_{A}
            + 
            \AB
        \right] 
        \left( 2\dot{r} n^j + v^j\right)
    \Bigg\}
    \,,
\end{align}
and from this we can read off the relative
PN order at which tidal effects affect the equation of motion for a binary system. As usual, we count every \textit{relative} factor of $GM/(rc^2)$ (as compared to the leading order contribution) as $1$PN order, 
and a power of $\dot{r}/c$ or $v^j/c$ as 1/2 PN order \cite{Blanchet:2013haa}.
From that counting, we see that the adiabatic finite-size correction, $\Lambda_{A}$,
enters as a 5 PN order correction to the point-particle term $(-GM/r^2) n^j$.
Similarly, the contribution from $\Xi_{A}$ appears as a 6.5 PN order correction
(see also \cite{Goldberger:2005cd}).
The fact that the leading-order finite size effect is 5 PN order smaller than the point particle
contribution is sometimes called the \textit{effacement principle}~\cite{1989thyg.book..128D}.

We next determine the 
leading-order conserved energy and the dissipation due to $\Xi_{A}$. First, we contract the center-of-mass acceleration in Eq.~\eqref{eq:eob-equations-of-motion-expanded}
with $\mu v^i$, and simplify to obtain  
\begin{align}
    &\frac{d}{dt} 
    \left[
        \frac{1}{2}\mu v_iv^i
        -
        \frac{G\mu M}{r}
        -
        \frac{3G^6 \mu M}{2c^{10}r^6}\left( m_B m_A^4
            \Lambda_{A}
            +
            \AB
        \right) 
    \right]
    \nonumber \\
    &= 
    -
    \frac{9G^7 \mu M}{c^{13}r^8}
    \left(
        m_B m_A^5\Xi_A
        +
        \AB 
    \right)
    \left(2\dot{r}^2 + v_iv^i\right)
    \,.
\end{align}
We reorganize this equation as
\begin{align}
\label{eq:newtonian-energy-dissipation}
    \frac{dE_{orb}}{dt}
    =&
    \mathcal{F}_{diss}\,,
\end{align}
where we have defined
\begin{subequations}
\begin{align}
    \label{eq:def-E-orb}
    E_{orb}
    \equiv&
    \frac{1}{2}\mu v_iv^i
    -
    \frac{G\mu M}{r}
    \nonumber\\
    &-
        \frac{3G^6 \mu M}{2c^{10}r^6}\left( m_B m_A^4
            \Lambda_{A}
            +
       m_A m_B^4
            \Lambda_{B}
        \right) ,
    \\
    \label{eq:def-F-vis}
    \mathcal{F}_{diss}
    \equiv&
    -
    \frac{9G^7 \mu M}{c^{13}r^8}
    \left(
        m_B m_A^5\Xi_A
        +
        m_A m_B^5\Xi_B
    \right)
    \nonumber\\
    &\times
    \left(2\dot{r}^2 + v_iv^i\right)
    \,.
\end{align}
\end{subequations}
The quantity $E_{orb}$ denotes the conserved orbital energy, which includes the contributions from the point-particle term and from $\lambda_{2,A}^{(0)}$. The energy flux due to tidal lag is represented by $\mathcal{F}_{diss}$.
As $\tau_{2,A}^{(1)}<0$ (see Eq.~\eqref{eq:motivation-dissipation-negative}), 
the orbital energy decreases in time due to tidal dissipative effects for non-spinning binaries. We emphasize that this correction is present even when one ignores gravitational waves, i.e., even when one does not back react the radiative losses in the system due to gravitational wave emission.

We next restrict the above equations to quasi-circular orbits.
Technically, circular orbit solutions to the above equations only exist when dissipation is set to zero. By ``quasi-circular orbits,'' we here mean those which deviate from circular orbits adiabatically, such that dissipation can be treated perturbatively. 
We therefore consider perturbative solutions~\cite{Flanagan:2007ix}, where both the conservative ($\Lambda_A$) and the dissipative ($\Xi_A$) tidal coefficients are assumed to introduce small deformations.
With this in mind, we use the solution ansatz 
\begin{subequations}
\label{eq:solution-x-ansatz}
\begin{align}
    x^i
    =&
    r(t)
    \left[\cos \varphi(t),\sin \varphi(t),0\right]
    ,\\
    r(t)
    =&
    r_0
    +
    \delta r(t)
    ,\\
    \varphi(t)
    =&
    \omega_0t 
    +
    \delta\varphi(t)
    .
\end{align}
\end{subequations}
Here $r_0$ and $\omega_0$ satisfy the equations of motion for a circular orbit, and thus, $\omega_0^2=GM/r_0^3$.
The perturbative quantities $\delta r$ and $\delta\varphi$ capture
corrections to circular orbit to linear order in $\Lambda_{A}$ and $\Xi_A$.
Inserting Eq.~\eqref{eq:solution-x-ansatz} into Eq.~\eqref{eq:eob-equations-of-motion-expanded},
we have
\begin{widetext}
\begin{align}
    \label{eq:linearized-newtonian-eom-quasi-circular}
    &
    \cos\left(\omega t\right)
    \left[
        9\left(\frac{c^4}{GM}\right)\frac{m_Bm_A^4}{M^5}\Lambda_{A}\gamma_0^7
        -
        3\left(\frac{c^3}{GM}\right)^2\gamma_0^3\delta r
        -
        2c\gamma_0^{1/2}\frac{d\delta\varphi}{dt}
        +
        \frac{d^2\delta r}{dt^2}
        +
        A\leftrightarrow B
    \right]
    \\
    &
    +
    \sin\left(\omega t\right)
    \left[
        -
        9\left(\frac{c^4}{GM}\right)\frac{m_Bm_A^5}{M^6}\Xi_{A}\gamma_0^{17/2}
        -
        2\left(\frac{c^3}{GM}\right)\gamma_0^{3/2}\frac{d\delta r}{dt}
        -
        \left(\frac{GM}{c^2}\right)\frac{1}{\gamma_0}
        \frac{d^2\delta\varphi}{dt^2}
        +
        A\leftrightarrow B
    \right]
    +
    \mathcal{O}\left(\delta r^2,\delta\varphi^2\right)
    =
    0
    \nonumber
    \,,
\end{align}
\end{widetext}
where
\begin{align}
     \gamma_0
    \equiv&
    \frac{GM}{r_0c^2}
    ,
\end{align}
Solving Eq.~\eqref{eq:linearized-newtonian-eom-quasi-circular} 
for $\delta r$ and $\delta\varphi$, 
we obtain the particular solutions
\begin{subequations}
\label{eq:perturbations-r-phi}
\begin{align}
    \delta r
    =&
    c_1
    +
    3\left(\frac{GM}{c^2}\right)\frac{m_Bm_A^4}{M^5}\Lambda_{A}\gamma_0^4
    \nonumber\\
    &
    -
    18\frac{m_Bm_A^5}{M^6} 
    \left(c t\right)\Xi_{A}\gamma_0^7 
    +
    A\leftrightarrow B
    ,\\
    \delta\varphi
    =&
    c_2t
    \nonumber\\
    &
    +
    \frac{27}{2}\frac{m_Bm_A^5}{M^6} 
    \left(\frac{c^3t}{GM}\right)^2
    \Xi_{A}\gamma_0^{19/2} 
    +
    A\leftrightarrow B
    ,
\end{align}
\end{subequations}
where $c_1$ and $c_2$ are constants of integration that must be related to each other by 
\begin{align}
        3\left(\frac{c^3}{GM}\right)^2\gamma_0^3c_1
        +
        2c\gamma_0^{1/2}c_2 = 0
\end{align}
for Eq.~\eqref{eq:linearized-newtonian-eom-quasi-circular} to be satisfied.  We must now make a choice about these constants of integration. 
We choose to set $c_1=c_2=0$, which sets $\omega\equiv d \varphi/dt = \omega_0$ at $t=0$.
Said another way, we choose the constants of integration such that $\omega_0$ is not just equal to the 
point-particle contribution to $\omega$, but is also the initial angular frequency. Note that this choice implies that 
$\gamma_0 \neq \gamma \equiv G M/(r_{\rm cons} c^2)$, where $r_{\rm cons}$ is the conservative part of the orbital separation. 
With our choices, we now have that these two $\gamma$'s are related by 
\begin{align}
    \gamma = \gamma_0 \left( 1 - 3 \frac{m_Bm_A^4}{M^5}\Lambda_{A}\gamma_0^5 \right)\,.
\end{align}

With these solutions in hand, we can now find the orbit-averaged energy and dissipative flux. 
We insert the conservative part of the solution to 
$r$ and $\varphi$ into the orbital energy expression in Eq.~\eqref{eq:def-E-orb},
and the tidal dissipative part into Eq.~\eqref{eq:def-F-vis}, to find 
\begin{subequations}
\begin{align}
\label{eq:orbit-average-expansion-E-small-viscosity}
    E_{orb}
    &=
    -
    \frac{1}{2}\mu c^2 \gamma_0 
    \nonumber \\
    &+
    \frac{9\gamma_0^6 \mu c^2}{2M^5}
        \left(m_Bm_A^4\Lambda_{A} 
        +
        m_Am_B^4\Lambda_{B}
        \right)
    ,\\
    \label{eq:Fdiss}
    \mathcal{F}_{diss}
    &=
    -
    \frac{9 \mu c^5 \gamma_0^9}{GM^7} 
    \left(m_Bm_A^5\Xi_{A} 
    +
    m_Am_B^5\Xi_{B} 
    \right)
    .
\end{align}
\end{subequations}

We define the orbital average of a quantity $A(t)$ as 
\begin{align}
    \stf{A} &= F \int_0^{\frac{1}{F}} dt A(t)\,,
\end{align}
where $F = \omega/(2 \pi)$ is the orbital frequency.
We can now orbit average the solutions given in Eq.~\eqref{eq:perturbations-r-phi} to obtain the secular changes in the orbital elements over one orbit
\begin{subequations}
\begin{align}
    \left\langle\frac{d \delta r}{dt}\right\rangle
    \approx&
    -
    18\frac{m_Bm_A^5}{M^6}c \Xi_{A}\gamma_0^7
    +
    A\leftrightarrow B
    ,\\
    \left\langle\frac{d\omega}{dt}\right\rangle
    \approx&
    27\frac{m_Bm_A^5}{M^6}\left(\frac{c^3}{GM}\right)^2\Xi_{A}\gamma_0^{19/2}
    +
    A\leftrightarrow B
    .
\end{align}
\end{subequations}
Tidal dissipation causes two initially non-spinning objects to
inspiral, $\left<dr/dt\right><0$ and $\left<d\omega/dt\right> >0$. Had the two object been rapidly
spinning, with a rotational frequency greater than the orbital frequency,
spin-orbit coupling would instead transfer spin angular momentum to the orbital
angular momentum, and we would have $\left<dr/dt\right>>0$ and $\left<d\omega/dt\right><0$ \cite{1981A&A....99..126H}.
\subsection{Approximate gravitational-wave phase}\label{sec:gw-phase}
Having computed the Newtonian equations of motion, 
we can now compute the phase of the Fourier transform of the gravitational waves emitted. Working in the stationary-phase approximation, we have that~\cite{Yunes:2009yz}
\begin{align}
    \label{eq:d2Psi-domega2-original}
    \Psi(f) = -2 \pi \int^{f/2} \tau' \left(2 - \frac{f}{F'}\right) dF'\,,
\end{align}
where $\tau' = F'/\dot{F}'$ and where the gravitational-wave frequency $f$ 
satisfies the stationary-phase condition $f=2 F$ (we consider an $\ell=2$ mode). 
This expression can be recast in a slightly simpler form.
Doing so, one finds that~\cite{Tichy:1999pv}
\begin{align}
    \label{eq:d2Psi-domega2}
    \frac{d^2\Psi}{df^2}
    =
    \frac{2\pi}{\dot{E}_{tot}}\frac{dE_{tot}}{df}
    \,.
\end{align}
For completeness, we provide a derivation of this result in Appendix~\ref{appendix:derivation-phasing}.
The total energy $E_{tot} = E_{orb}$ is given in Eq.~\eqref{eq:orbit-average-expansion-E-small-viscosity}. 
In the adiabatic approximation, the rate of change of the total energy can be related to the total energy flux via the balance law
\begin{align}
    \label{eq:energy-balance-general}
    \frac{d}{dt}E_{tot}
    =
    \mathcal{F}_{tot}
    ,
\end{align}
For the binaries we consider, the total energy flux is nothing but
\begin{align}
\label{eq:etot-ftot}
    \mathcal{F}_{tot}
    =&
    \mathcal{F}_{GW}
    +
    \mathcal{F}_{diss}
    ,
\end{align}
where $\mathcal{F}_{GW}$ is the energy flux due to gravitational wave emission and $\mathcal{F}_{diss}$ is the energy flux due to tidal dissipation, which we derived in Eq.~\eqref{eq:Fdiss}. 

We use the quadrupole formula to determine the leading-order energy flux due to gravitational waves 
(see \cite{Poisson-Will} for a recent review). To leading PN order, we have
\begin{align}
    \mathcal{F}_{GW}
    =
    -
    \frac{G}{5 c^5}
    \left<
        \frac{d^3I_{\stf{ij}}^T}{dt^3}
        \frac{d^3I_{\stf{ij}}^T}{dt^3}
    \right>
    ,
\end{align}
where the total conservative quadrupole moment, $I^T_{ij}$, is
(recall there is no summation of capital Latin indices)
\begin{align}
    I_{ij}^T
    &=
    I^A_{ij} + I^B_{ij}
    +
    \mu x_ix_j \,,
    \nonumber\\
    &= 
    -
    \frac{G^4}{c^{10}}\left(\Lambda_{A} m_A^5\mathcal{E}_{ij}^A +\AB \right)
    +
    \mu x_i x_j
    .
\end{align}
Making use of Eqs.~\eqref{eq:quadrupole-moment-Iij},
\eqref{eq:solution-x-ansatz}, and \eqref{eq:perturbations-r-phi}, 
and expanding to linear order in $\Lambda_A$, we find that
(c.f. \cite{Flanagan:2007ix})
\begin{align}
    \label{eq:g-gw-gamma}
    &\mathcal{F}_{GW}
    \approx
    -
    \frac{32}{5}\frac{c^5}{G}\eta^2 \gamma_0^5
    \Bigg[ 
        1
        \nonumber\\
        &
        \hspace{0.5cm}
        +
        6
        \left(1+3\frac{m_B}{m_A}\right)\left(\frac{m_A}{M}\right)^5\Lambda_{A}
        \gamma_0^5
        +
        A\leftrightarrow B
    \Bigg]
    .
\end{align}
We note that we have only included the adiabatic tidal correction in calculating
$\mathcal{F}_{GW}$; the contribution of $\Xi_A$ is
captured entirely in the term $\mathcal{F}_{diss}$.

We now have all the tools needed to compute the Fourier phase. Using Eq.~\eqref{eq:d2Psi-domega2} and PN expanding, we find
\begin{align}
\label{eq:stationary-phase-approximation-viscosity-d2}
    \frac{d^2\Psi}{df^2}
    &=
    \frac{5\pi^2}{48}\frac{1}{\eta}\left(\frac{GM}{c^3}\right)^2 u^{-11}
    \Bigg[
        1
        \nonumber \\
        &-
        \frac{45}{32}\frac{1}{\eta}\frac{m_Bm_A^5}{M^6}\Xi_{A} u^8
        -
        6\left(1 + 12\frac{m_B}{m_A}\right)\left(\frac{m_A}{M}\right)^5
        \Lambda_{A} u^{10}
        \nonumber \\
        &+
        A\leftrightarrow B
    \Bigg]
    .
\end{align}
where we have introduced 
\begin{align}
    \label{eq:udef}
    u
    \equiv
    \left(\frac{G \pi M f}{c^3}\right)^{1/3}
    ,
\end{align}
and we have used that $\gamma_0 = u^2$ in the stationary-phase approximation.
Integrating this twice and using Eq.~\eqref{eq:lambdabar-xibar}, we find
\begin{align}
\label{eq:stationary-phase-approximation-viscosity}
    \Psi\left(f\right)
    &=
    \frac{3}{128}\frac{1}{\eta} u^{-5}
    \left[
        1
        -
        \frac{75}{32}
        \bar{\Xi} \,
        u^8 \log(u)
        -\frac{39}{2}\bar{\Lambda}u^{10}
    \right]
    \nonumber \\
    &+
    2\pi f \bar{t}_c
    -
    \varphi_c
    -
    \frac{\pi}{4}
    \,,
\end{align}
where we have redefined the coalescence time via 
\begin{align}
    \bar{t}_c = t_c + \frac{75 G M \bar{\Xi}}{8192
    c^3 \eta }\,.
\end{align}
We see that $\bar{\Xi}$ appears at 4PN relative order in the phase,
and that the presence of $\mathrm{log}\left(u\right)$ ensures that it is
not degenerate with the time of coalescence. 
Why do these effects enter the gravitational-wave phase at 4PN order, when out-of-equilibrium effects enter the orbital dynamics at 6.5PN order
(see Eq.~\eqref{eq:eob-equations-of-motion-expanded} and Sec.~\eqref{sec:effective-point-particle})? The key to the answer of this question lies in Eq.~\eqref{eq:etot-ftot}. Within the adiabatic
approximation, the gravitational-wave phase depends on \textit{both} the derivative of the
quasi-adiabatic orbital energy $E_{orb}$ and on the dissipation $\mathcal{F}_{diss}$.
The adiabatic tidal Love numbers only enter $E$, while, by contrast, out-of-equilibrium effects enter $\mathcal{F}_{diss}$.

The above conclusion, that viscous effects enter gravitational wave observables at 4PN order, was first speculated about in~\cite{Most:2021zvc} through a ``Fermi'' (order-of-magnitude) estimate. 
The authors there found a 4PN viscous correction to the stress-energy tensor of a neutron star. 
Using this, the authors estimated that such modifications would also enter gravitational waves at 4PN order. 
While dissipative corrections do enter the gravitational wave phase at 4PN, the correction computed in \cite{Most:2021zvc} enter the phase at higher than 4PN order. This is because corrections to the stress-energy tensor only provide \emph{local} information in the rest frame of the object in the binary system, and the microscopic properties of a star (through their effects on the star's multipole moments) enter at higher PN order in the star's equations of motion \cite{1989thyg.book..128D}.
We expect that lower PN order corrections to a neutron star's stress-energy tensor will largely determine the dissipative tidal deformability. 
Finally, we note that the estimate in \cite{Most:2021zvc} additionally only found a finite-size enhancement of 
compactness $C^{-2}$ to the gravitational wave phase, while we find a finite-size enhancement of
compactness $C^{-6}$ (see Eq.~\eqref{eq:definition-tidal-deformabilities} and Eq.~\eqref{eq:stationary-phase-approximation-viscosity}).
\section{Regime of applicability and detection plausibility\label{sec:regime-of-applicability}}

Here we discuss the regime of applicability of our derivation of the phasing formula Eq.~\eqref{eq:stationary-phase-approximation-viscosity}. 
We additionally outline how a constraint on the dissipative tidal deformability $\Xi_A$ could be mapped to a constraint on the effective micro-physical properties of the star, namely the effective bulk and shear viscosities of the neutron star fluid.

To estimates the values $\Xi$ can take one needs to calculate the value of $k_2$ and $\tau_d$ for different stellar models.
These have been computed for (Schwarzschild) black holes~\cite{Poisson:2020vap} and giant stars~\cite{Ogilvie_2014}
.
For a Schwarzschild black hole of mass $M$, $k_2=0$ and 
\begin{align}
    k_2 \tau_d &= \frac{G M}{30 c^3} \,.
\end{align}
For giant stars the value of $\tau_d$ depends on the detailed micro-physical properties of the stellar model~\cite{Ogilvie_2014,Zahn:2008fk}.
A common parameterization of the tidal lag time for the linear tidal response, that we adopt here, is
\begin{align}
    \label{eq:tidal-lag-def}
    \tau_{2,A}^{(1)}
    \equiv
    -
    \tau_{d,A} 
    \equiv 
    -
    \frac{p_{2,A} \nu_A R_A}{G m_A}
    ,
\end{align}
where $p_{2,A}$ is a dimensionless number and $\nu_A$ is an averaged quantity which has the dimensions of kinematic viscosity 
\begin{align}
    \label{eq:def-kinematic-viscosity}
    \nu
    \equiv 
    \frac{\left<\eta\right>}{\left<\rho\right>}
    .
\end{align}
Here $\left<\eta\right>$ is a volume average of the shear and bulk viscosities of the fluid, and $\left<\rho\right>$ is the average density of the star. We emphasize that Eq.~\eqref{eq:tidal-lag-def} is a phenomenological parametrization of $\tau_{2,A}^{(1)}$, and that the coefficient $p_{2,A}$ is currently unknown for 
realistic equation of state (EOS).
For relativistic polytropes\footnote{We derive these quantities in an upcoming work~\cite{future-work-viscosity}.}, the values of $p_2$ for bulk viscous flow lie in the range $0.02 - 0.2$ and for shear viscous flow lie in the range $0.1-10$.
For the purposes of this paper, we pick $p_2 = 0.1$ for bulk viscous flow and $p_2 = 5$ for shear viscous flow.
We emphasize that the invariant quantity that appears in the waveform is the $\tau_d$ and we use these values for $p_2$ to see how large $\tau_d$ can be for a given value of shear/bulk viscosity.

With the parametrization of Eq.~\eqref{eq:tidal-lag-def}, we now determine the regime of
applicability of the assumptions we made in deriving the gravitational-wave phase formula
of Eq.~\eqref{eq:stationary-phase-approximation-viscosity}.
In particular, we assumed that the neutron stars were not tidally locked.
We first revisit and confirm the old result that tidal locking would require unphysically
large values of the viscosity.
For the sake of completeness we then show that, if, for whatever reason, the bodies were
tidally locked, viscosity would affect the gravitational-wave phase at 2PN order.
This being said, one could only conclude a \emph{lower} bound for the viscosity
if the stars are tidally locked, as the effective viscosity does not explicitly
enter the tidal locking terms in the gravitational wave phase. 

We next demonstrate that the dissipative timescale, orbital timescale of the binary,
and the characteristic internal timescale of neutron stars satisfy the inequalities 
given in Eq.~\eqref{eq:tower-time-inequalities} for values of viscosity that also satisfy
the no-tidal-locking condition. We conclude that our expansion of the
quadrupole to linear order in $d/dt$, Eq.~\eqref{eq:newtonian-first-order-limit-Iij}, 
should be accurate\footnote{Higher order terms in
$d/dt$ in the expansion of Eq.~\eqref{eq:newtonian-first-order-limit-Iij}
may be required to accurately determine the tidal Love number; for more
discussion see \cite{Pratten:2021pro})}.

Finally, we show that physically allowed values of $\stf{\zeta},\stf{\eta}$ ($\tau_d$) could lead to an amount of dephasing comparable, and even larger than, the dephasing caused by the adiabatic tidal Love number $k_2$.
We use a Fisher analysis to show that there is a window of parameter space for physically allowable
values of $\tau_d$ that could be constrained with future gravitational-wave observations,
although more work needs to be done to obtain precise theoretical estimates for
$\tau_d$ realistic EOS.
Finally, in this section we show that for values of the tidal lag that are larger than $\tau_d$ > $21 \mu\mathrm{s}$, the effective dissipative tidal deformability could be measurable with current ground-based gravitational detectors.

Before continuing, we discuss the various estimates for $\nu$ that have been proposed over time. 
In \cite{Most:2021zvc}, the effective bulk viscosity due to Urca processes in the star is reported to be as large as  $\zeta/\rho\sim 10^{15}\mathrm{cm}^2/\mathrm{s}$ during the late inspiral (here $\rho$ is the rest-mass energy density)\footnote{More recent nuclear physics calculations suggest $\zeta/\rho$ could locally reach values as high as $10^{17}\mathrm{cm}^2/\mathrm{s}$\cite{Yang:2023ogo}.}.
This value is much larger than the effective kinematic viscosity predicted by microscopic calculations of the shear viscosity of neutron stars, current estimates of which range from $\nu\sim 10^{4-6}\mathrm{cm}^2/\mathrm{s}$ \cite{Shternin:2008es}.\footnote{We note that even if one consider earlier estimates of the molecular shear viscosity of $\nu \lesssim 10^9\mathrm{cm}^2/\mathrm{s}$ \cite{1979ApJ...230..847F}, the upper end of this range is still likely too small to make any measurable impact on the gravitational wave phase}.

The molecular bulk viscosity of the star will likely be many orders of magnitude larger than the molecular shear viscosity of the star.
Nevertheless, in principle there could sizeable contributions to the neutron star shear viscosity through the crust or ``anomalous'' viscosities. 
Estimates of the shear viscosity of the neutron star crust \cite{1992ApJ...398..234K,Shternin:2008es} range over $\nu\sim 10^{6-15}\mathrm{cm}^2/\mathrm{s}$\footnote{We caution that in effect the upper bound $\nu\sim 10^{15} \mathrm{cm}^2/\mathrm{s}$ quoted in \cite{1992ApJ...398..234K} is a theoretical in-principle upper bound for how large the shear viscosity contribution of the crust could be; we are unaware of detailed calculations of the crust effective shear viscosity that give values this large.}.
The turbulent ``anomalous'' viscosities of neutron stars remain much less well understood, although they play an important role in viscous properties of main-sequence stars \cite{1977A&A....57..383Z}. 
Determining the contribution of turbulent effects will require a more detailed understanding of the interaction of driven mode solutions to the neutron star, and is outside the scope of this work. 
\subsection{Tidal torquing,
and the regime of applicability of the no-spin approximation\label{sec:regime-applicability}}

In our solution to the Newtonian equations of motion, 
we neglected the dynamical effects of the tidal torquing of the star
due to the misalignment of the stars' quadrupole moment and the external
gravitational field.
Such tidal torquing transfers orbital angular momentum to the stars, which
in turn affects the orbital dynamics.
For a review of these concepts in the purely Newtonian context, see \cite{Poisson-Will}.
Bildsten and Cutler \cite{1992ApJ...400..175B} estimated the effects of tidal
torquing and locking on the dynamics of compact binaries, 
and concluded that the effects would be marginal 
because neutron stars could not tidally lock before merger.
Here we revisit their calculation, 
and estimate the PN order at which the tidal spin up enters the gravitational-wave phase. Ultimately we reach the same conclusion as those authors, i.e.~that the viscosity
required to achieve tidal locking before merger would exceed physically reasonable values of that coupling.

We assume that object A is spinning with spin vector $S_A^i = e_A^i S_A$,
where $e_A^i$ is a unit vector and $S_A$ is the magnitude of the spin vector.
In the rest frame of object $A$, the magnitude of the spin vector
obeys the following evolution equation \cite{Poisson-Will}
\begin{align}
    \label{eq:general-spin-evolution}
    \frac{dS_A}{dt}
    =&
    -
    \epsilon_{ijk}e^iI_A^{\stf{jp}}\left(\mathcal{E}_A\right)^k_p
    .
\end{align}
Using Eq.~\eqref{eq:tidal-field-EA} for $\mathcal{E}_A^{ij}$
and Eq.~\eqref{eq:quadrupole-moment-Iij} for $I_A^{ij}$, we find
\begin{align}
    \frac{dS_A}{dt}
    =
    -
    \frac{9G^7m_A^6m_B^2}{c^{13}r^8}
    \Xi_{A}
    \epsilon_{ijk}e^iv^jx^k
    .
\end{align}
We further assume that the stars are initially not spinning and use Eq.~\eqref{eq:solution-x-ansatz} to set
$e_A^i
    =
    \left(0,0,1\right)
    $. 
Note that as tidal torquing spins up the stars in the orbital
plane, the orbit remains circular and there is no precession
(for more discussion and reviews, see
\cite{Barker:1975ae,1989thyg.book..128D,Poisson-Will}).

Using Eq.~\eqref{eq:solution-x-ansatz} for $x^i$,
and working to linear order in $\Xi_A$, we find that
\begin{align}
    \frac{dS_A}{dt}
    \approx&
    \frac{9Gm_B^2 m_A^6}{M^6c}\Xi_{A}\gamma_0^6\Delta\omega_A 
    ,
\end{align}
where we defined $\Delta\omega_A\equiv\omega_0-\Omega_A$ 
to be the difference in the orbital frequency 
$\omega_0$ and the star's rotational frequency $\Omega_A$. 
If we set $S_A \approx (2/5)m_A R_A^2 \Omega_A$, 
($I\approx (2/5)m_AR_A^2$ is the approximate moment of inertia of the star), 
we obtain a differential equation for $\Omega_A$ (compare to \cite{1981A&A....99..126H})
\begin{align}
    \label{eq:change-in-particle-freq}
    \frac{d\Omega_A}{dt}
    \approx&
    \frac{45Gm_B^2 m_A^5}{2R_A^2 M^6c}\Xi_{A}\gamma_0^6
    \left(\omega_0-\Omega_A\right)
    . 
\end{align}
This gives us a characteristic tidal torquing timescale of 
\begin{align}
    &T_{A}
    \equiv
    \frac{2 R_A^2 M^6 c}{45 G m_B^2 m_A^5}
    \frac{1}{\Xi_{A}}\gamma_0^{-6}
    \nonumber\,,\\
    &=
    \frac{1}{15}\frac{G M^6}{m_B^2 m_A^3 c^4}
    \frac{R_A}{\tau_{d,A}}
    \frac{C_{A}^4}{k_{2,A}}
    \gamma_0^{-6}
    \nonumber \,, \\
    &\approx
    2.3 \times 10^3 \mathrm{s}
    \left(\frac{M}{3.2 \rm M_{\odot}}\right)^6
    \left(\frac{1.6 \rm M_{\odot}}{m_B} \right)^2
    \left(\frac{1.6 \rm M_{\odot}}{m_A} \right)^2
    \left( \frac{0.1}{k_{2,A}}\right)
    \nonumber \\
    &\times 
    \left(\frac{C_A}{0.196}\right)^4
    \left( \frac{0.1}{p_{2,A}}\right)
    \left( \frac{10^{14}\, \rm cm^2 s^{-1}}{\left<\nu\right>}\right) \left(\frac{0.072}{\gamma_0} \right)^{6}
    .
\end{align}
On the second line above, we have recast the characteristic tidal torquing timescale in terms of the tidal Love number, and the tidal dissipation timescale 
$\tau_{d,A}=-\tau^{(1)}_{2,A}$.

If the tidal torquing timescale is smaller than the time it takes the binary to inspiral, then there is a chance for tidal torquing to occur. 
To determine the leading gravitational-wave phase effect due to tidal torquing, 
we only need to consider the dissipation due to the emission of gravitational waves. 
In this case, the orbital radius changes approximately as \cite{Peters:1964zz}
\begin{align}
    \frac{dr}{dt}
    \approx
    -
    \frac{64}{5}\eta \frac{G^3M^3}{c^5r^3}
    \,,
\end{align}
which allows us to define the characteristic inspiral timescale 
\begin{align}
    T_{insp}
    &\equiv
    \frac{5}{64}\frac{1}{\eta}\frac{GM}{c^3}\gamma_0^{-4}\,,
    \nonumber \\
    &=
    1.7 \times 10^{-1}
    \left(\frac{0.25}{\eta}\right)
    \left( \frac{M}{3.6 \rm M_{\odot}}\right)
    \left( \frac{0.07}{\gamma_0}\right)^{4}
    .
\end{align}
The ratio of the tidal locking time to the inspiral time is
\begin{align}
    \frac{T_A}{T_{insp}}
    &=
    \frac{64 }{75}
    \frac{ G M^3 }{c m_B m_A\left<\nu_A\right>}
    \frac{C_{A}^4}{k_{2,A} p_{2,A}}
    \gamma_0^{-2}
    \nonumber \\
    &\approx
    1.33 \times 10^{4}
    \left( \frac{M}{3.2 \mathrm{M}_{\odot}}\right)^3
    \left(\frac{1.6 \mathrm{M}_{\odot}}{m_B}\right)
    \left(\frac{1.6 \mathrm{M}_{\odot}}{m_A}\right)
    \nonumber \\
    & \left( \frac{10^{14} \mathrm{cm}^2 \mathrm{s}^{-1} }{\left<\nu_A\right>}\right)
    \left( \frac{0.1}{p_{2,A}}\right)
    \left( \frac{0.1}{k_{2,A}}\right)
    \left(
    \frac{0.07}{\gamma_0}
    \right)^2
    \left( 
    \frac{C_A}{0.2}
    \right)^4\,.
    \label{eq:ratio_TA_insp}
\end{align}
When the torquing and inspiral timescales are comparable, $T_A/T_{insp}\sim1$,
the stars have enough time to, in principle, become tidally locked. We see, however, that this is not the case for typical values of the tidal Love number and the $p_{2,A}$ dissipation coefficient for either the shear viscous case $p_2 \sim 5$ or the bulk viscous case shown above.

We emphasize that Eq~\eqref{eq:ratio_TA_insp}, while derived using a standard
set of assumptions \cite{1992ApJ...400..175B}, 
provides only a rough measure of when two stars could become tidally locked. 
The quantities $T_A$ and $T_{insp}$ are characteristic timescales; we may expect that
the actual tidal locking and inspiral times will be longer than $T_A$ and $T_{insp}$
by factors that we have not computed. For example, we have fixed $\gamma_0$ to be a constant
reference value (this reference value corresponds to an orbital frequency of $F=400$Hz,
with $M=3.2\mathrm{M}_{\odot}$), 
even thought it varies with time as the binary inspirals to merger. 
A self-consistent calculation, that does not hold
$\gamma_0$ fixed, and that takes into account the change in the orbital frequency
as angular momentum is transferred to the two stars, is necessary to fully
determine whether or not tidal locking could feasibly happen before merger.

To understand what the maximum, physically reasonable value that $\left<\nu\right>$ could be (given the parameterization Eq.~\eqref{eq:tidal-lag-def}), we define the maximum viscosity that is consistent with a volume averaged notion of causal momentum transport across the star. 
We do so by realizing that the rate of momentum diffusion can be no larger than the speed of light for a fluid to respect causality. 
This then implies that
\begin{align}
    \label{eq:max-value-kinematic-viscosity}
    \left<\nu_{\rm causal}\right> \equiv c \, R 
    =
    3.6\times 10^{16}\mathrm{cm}^2\mathrm{s}^{-1}
    \left(\frac{R}{12\mathrm{km}}\right).
\end{align}
That is, volume-averaged, effective kinematic viscosities that approach this value imply momentum diffusion across the star that occur at around the speed of light. 
This is a larger value than the (widely varying) estimates for the effective kinematic viscosity of neutron stars. 
Given this, and that $\gamma_0\ll1$, we conclude that neutron star binaries cannot tidally synchronize before merger without exceed the limit Eq.~\eqref{eq:max-value-kinematic-viscosity} \cite{1992ApJ...400..175B} (see also \cite{1992ApJ...398..234K} for a related discussion).

Finally, for the sake of completeness, we estimate the PN order at which tidal locking would
enter the gravitational waveform if it were to be realized in nature.
From Eq.~\eqref{eq:ratio_TA_insp}
tidal locking in principle could be important if the stars were sufficiently non-compact, because the timescale decays with the fourth power of compactness.
For a tidally-locked star, the rotational frequency is equal to the orbital frequency, 
so that its spin vector is
\begin{align}
    S_A
    \approx&
    I_A \omega
    ,
\end{align}
where $I_A$ is the moment of inertia of the star.
We consider a quasi-circular orbit of radius 
$r_0$ and orbital frequency $\omega_0$.
The approximate Newtonian energy then is
\begin{align}
    E_{orb}
    &=
    \frac{1}{2}\mu v_iv^i
    -
    \frac{G\mu M}{r}
    +
    \frac{1}{2}I_A\omega_0^2
    \nonumber\\
    &
    =
    -
    \frac{1}{2}\mu c^2 \gamma_0
    \left(
        1
        -
        \frac{2}{5}\frac{m_A^2}{Mm_B}
        \frac{1}{C_A^2}
        \gamma_0^2
    \right)
    \,,
\end{align}
where again, as a leading-order estimate, we have approximated the neutron star
moment of inertia as $I_A\approx (2/5)R_A^2m_A$.
We set the energy flux to be the leading-order contribution
from gravitational waves 
\begin{align}
    \mathcal{F}_{tot}
    =
    -
    \frac{32}{5}\frac{c^5}{G}\eta^2\gamma_0^5
    .
\end{align}
We then set $\gamma_0=u^2$, 
integrate Eq.~\eqref{eq:d2Psi-domega2} twice, and find that 
\begin{align}
    \Psi\left(f\right)
    \approx&
    \frac{3}{128}\frac{1}{\eta}u^{-5}
    \left[
        1
        -
        12\frac{m_A^2}{Mm_B}\frac{1}{C_A^2}u^4
    \right]
    \nonumber\\
    &
    +
    2\pi f t_c - \Psi_c - \frac{\pi}{4}
    .
\end{align}
We see that tidal locking enters at 2PN relative order in the gravitational-wave phase\footnote{We note this
is $0.5$PN lower than the relative $2.5$PN order at which dissipative effects enter for black hole
binaries with arbitrary spin \cite{Poisson:1994yf,Tagoshi:1997jy,Alvi:2001mx,Porto:2007qi}.}.
This being said, we emphasize that unphysically large values
of viscosity are required for neutron star binaries to be tidally locked. 
Moreover, even if the effects of tidal locking could be measured,
it would provide only a \emph{lower} bound on the viscosity of the star,
as the viscosity does not explicitly enter the gravitational-wave phase at this PN order.

\subsection{Applicability of our first-order truncation of the tidal quadrupole}\label{sec:inequality-verification}

We now verify that the inequalities given in Eq.~\eqref{eq:tower-time-inequalities}
are satisfied for physically reasonable values of binary neutron star parameters. 
We emphasize that in this section we assume that out-of-equilibrium effects 
are driven primarily by the effective kinematic viscosity of the star (see, Eq.~\eqref{eq:tidal-lag-def}). 
We set the stars' internal timescale to be
$\tau_{int} = \sqrt{R_A^3/(GM_A)}$, which roughly corresponds to the
period of the $f-$mode of a neutron star \cite{Kokkotas:1999bd},
set $\tau_{sys}=2/f$, the orbital period of the binary, and, as before, we
set $\tau_d=p_{2,A}\left<\nu_A\right>R_A/(Gm_A)$ (see Eq.~\eqref{eq:tidal-lag-def}).
Inserting these into Eq.~\eqref{eq:tower-time-inequalities}, we find
\begin{widetext}
\begin{subequations}
  \begin{align}
    \frac{\tau_d}{\tau_{sys}}
    &= 
     \frac{p_{2,A} \left<\nu_A\right> f }{2 c^2 C_A} 
    =6.27 \times 10^{-4}
    \left(\frac{p_{2,A}}{0.1}\right)
    \left(\frac{\left<\nu_A\right>}{10^{14} \mathrm{cm}^2 \mathrm{s}^{-1}} \right)
    \left(\frac{0.2}{C_A} \right)
    \left(\frac{f}{400 \mathrm{Hz}} \right) \,, \\
    \frac{\tau_{int}}{\tau_{sys}}
    &=
    \frac{G m_A f}{2 c^3 C_A^{3/2}}
    = 
    1.8 \times 10^{-2}
    \left( \frac{m_A}{1.6\mathrm{M}_{\odot}}\right)
    \left(\frac{f}{400 \mathrm{Hz}}\right)
    \left(\frac{0.2}{C_A} \right)^{3/2}\,.
\end{align}  
\end{subequations}
\end{widetext}
Comparing the above estimates with Eq.~\eqref{eq:tower-time-inequalities},
we see that when $p_{2,A} \left<\nu_A\right>\gtrsim 10^{18}\mathrm{cm}^2\mathrm{s}^{-1}$, 
our approximation starts to break down.
We see that this is also in the regime in which the tidal locking time scale starts approaching 
the time scale of inspiral of Eq.~\eqref{eq:ratio_TA_insp}.
Given the unphysically large values of viscosity required for tidal locking, 
we see from the above equations that our approximation can be used safely to 
describe the linear response of the quadrupole to the external field in the inspiral.

\subsection{Detection plausibility\label{sec:detection-plausibility}}
To estimate the observational relevance of the viscosity for gravitational-wave observations, we first estimate the number of radians of ``dephasing'' that viscous
corrections could introduce to the gravitational-wave phase.
From Eqs.~\eqref{eq:definition-tidal-deformabilities},
\eqref{eq:stationary-phase-approximation-viscosity}, and \eqref{eq:tidal-lag-def}, we have
\begin{align}
    \Delta\Psi_{\Xi}
    &
    \approx
    -\frac{75 c}{256 G M^2  }
    \left(
    \frac{m_A^2 k_{2,A} \left<\nu_A\right> p_{2,A}}{m_BC_A^6}
    +
    \AB
    \right)
    \nonumber \\
    &\left( 
    u_f^{3} 
    \log\left( u_f\right)
    +\FI
    \right)
\end{align}
where the subscripts $i/f$ stand for initial/final values.
As before, as a reference we consider an equal-mass neutron star binary, $m_A=m_B$ and set $f_i = 0$ to obtain the maximum dephasing
\begin{widetext}
\begin{align}
    \label{eq:Delta-Psi-Lambda1}
    \Delta\Psi_{\Xi}
    \approx&
    -9.41 \times 10^{-3}
    \left(
        \frac{3.2 \mathrm{M}_{\odot}}{M}
    \right)^2
    \left[ 
        \left(\frac{m_A}{1.6 \mathrm{M}_{\odot}}\right)^2 
        \left(\frac{1.6 \mathrm{M}_{\odot}}{m_B}\right)
        \left(\frac{\left<\nu_A\right>}{10^{14} \mathrm{cm}^2 \mathrm{s}^{-1}} \right)
        \left(\frac{k_{2,A}}{0.1}\right)
        \left(\frac{p_{2,A}}{0.1}\right)
        \left(\frac{0.2}{C_A}\right)^6
        +
        \AB
    \right]\nonumber\\
    &
    \times 
    \left[
        \left(\frac{f}{400\mathrm{Hz}} \frac{M}{3.2 \mathrm{M}_{\odot}}\right)
        \log\left(\frac{f}{400\mathrm{Hz} }\frac{M}{3.2 \mathrm{M}_{\odot}}\right) 
    \right]
    .
\end{align} 
As a point of comparison, we can similarly estimate the dephasing due to tidal Love numbers to be
\cite{Flanagan:2007ix}
\begin{align}
    \label{eq:Delta-Psi-Lambda0}
    \Delta\Psi_{\Lambda}
    \approx&
    -
    \frac{3}{8 \eta M^5}
    \left(
        \frac{m_A^5 k_{2,A}}{C_A^5}\left(1 + \frac{12 m_B}{m_A}
    \right)
    +\AB
    \right)
    \left(u_f^{5} - \FI \right)
    \nonumber\\
    \approx&
    -
    0.6
    \left(\frac{0.25}{\eta}\right)
    \left(\frac{3.2 \mathrm{M}_{\odot}}{M} \right)^5
    \left[
        \left(\frac{m_A}{1.6 \mathrm{M}_{\odot}}\right)^5
        \left(\frac{k_{2,A}}{0.1}\right)
        \left(\frac{0.2}{C_A}\right)^5
        \left(\frac{1 + 12 m_B/m_A}{13}\right)
        +
        \AB
    \right]
    \nonumber\\
    &
    \times
    \left[
        \left( \frac{f}{400\mathrm{Hz}} \frac{M}{3.2 \mathrm{M}_{\odot}}\right)^{5/2}
    \right]
    .
\end{align}
\end{widetext}
The dephasing due to bulk viscous effects is smaller than the dephasing due to the adiabatic tide.
If one chooses $p_2 \sim 5$ corresponding to the shear viscous case then we see that $\Delta \Psi_{\Xi}  \sim -0.47$ and we see that this dephasing is comparable to the case of the adiabatic tide.

Dephasing is not the full story, however, since parameter degeneracies can greatly deteriorate our ability to extract new physics from the waveform, especially when this new physics enter at high PN order. Therefore, to better estimate the measurability of $\Bar{\Xi}$, we now perform a simple Fisher analysis~\cite{Finn,Cutler-Flanaggan}, which should provide an upper bound on measurement accuracy.
We first quickly summarize the basics of a Fisher analysis.
Consider a signal model paramtrized by parameters $\Theta^a$, where the Latin index ranges over the waveform parameters only.
The (square of the) signal-to-noise ratio (SNR) is defined as
\begin{equation}
    \rho^2 \equiv 4 \int_{f_{\rm low}}^{f_{\rm upper}} \frac{ \tilde{h}(f)\tilde{h}^{*}(f)}{S_{n}(f)}\,df,
\end{equation}
where $\tilde{h}(f)$ is the frequency domain waveform, a superscript star stands for complex conjugation, and $S_n(f)$ is the (one-sided) noise power spectral density (PSD) of the detector. 
We set $f_{low} = 10\, \mathrm{Hz}$ and $f_{upper} = \mathrm{min} \left(f_{\rm ISCO}, f_{\rm cont} \right)$, where
$f_{\rm ISCO} = c^3/(G 6^{3/2} \pi M)$ is the frequency at the innermost stable circular orbit of a Schwarzschild
black hole of mass $M$, and $f_{\rm cont} = \sqrt{G M/(R_1+R_2)^3}/\pi$ is the frequency at the point
of contact of the two stars.
In the limit of large SNR and Gaussian, stationary noise, 
the probability that the gravitational-wave data $d(t)$ is characterized by the source parameters $\Theta^a$ is approximately
\begin{equation}\label{posterior}
    p(\Theta |d) \approx p^{0}(\Theta) \exp\Big[-\frac{1}{2}\Gamma_{ab}(\Theta^{a} - \hat{\Theta}^{a})(\Theta^{b} - \hat{\Theta}^{b})\Big],
\end{equation}
where $p^{0}(\Theta)$ is the prior probability, and $\hat{\Theta}^{a}$ is the maximum of the Gaussian likelihood function. The Fisher information matrix $\Gamma_{ab}$ is defined as
\begin{align}
    \Gamma_{ab} 
    &= 
    2 \int_{0}^{\infty} \frac{df}{S_{n}(f)}\left( 
        \frac{\partial \tilde{h}(f)}{\partial \Theta_a}\frac{\partial \tilde{h}^*(f)}{\partial \Theta_b}
        + 
        \frac{\partial \tilde{h}(f)}{\partial \Theta_b}\frac{\partial \tilde{h}^*(f)}{\partial \Theta_a}
    \right) 
    .
\end{align}
The inverse of the Fisher matrix is the variance-covariance matrix, 
\begin{equation}
    \Sigma_{ab} \equiv \Gamma_{ab}^{-1}\,,
\end{equation}
whose diagonal elements provide the square of the 1$\sigma$ errors on the estimation of waveform parameters
(no summation on repeated indices here),
\begin{equation}
\Delta \Theta_{a} = \sqrt{\Sigma_{aa}} .
\end{equation}
If one uses assumes a Gaussian prior with a width of $\sigma_{\Theta}$ in the analysis, then the $\Gamma_{ab}$ in the above formula is replaced by
\begin{align}
    \Tilde{\Gamma}_{ab} &= \Gamma_{ab} + \frac{1}{\sigma_{\Theta}^2} \delta_{ab}\,.
\end{align}

For our problem, the Fourier domain gravitational waveform can be modeled in the restricted post-Newtonian approximation via
\begin{align}
    \Tilde{h}(f)
    &=
    \mathcal{A} f^{-7/6} e^{i \Psi (f)}\,.
\end{align}
The phase can be separated into a point-particle piece, an adiabatic tidal piece and a dissipative tidal piece [Eq.~\eqref{eq:stationary-phase-approximation-viscosity}]
\begin{align}
    \Psi &= \Psi_{pp} + \Psi_{\Lambda} + \Psi_{\Xi}\,.
\end{align}
We include corrections up to 4.5 PN order for the point particle piece~\cite{blanchet2023gravitationalwave}
\begin{align}\label{eq:psi-pp}
    \Psi_{pp}
    &=
    -\frac{\pi}{4}
    -
    \varphi_c
    +
    2 \pi f t_c
    +
    \frac{3}{128 \eta u^{5}}
    \left[ 
    \sum_{k=0}^{9}
    \alpha_k u^{k}
    \right]
    \,,
\end{align}
where the coefficients $\alpha_k$ are given in~\cite{blanchet2023gravitationalwave}
.
The adiabatic tidal contribution and the dissipative tidal contributions are given respectively by [Eq.~\eqref{eq:stationary-phase-approximation-viscosity}]
\begin{subequations}
\begin{align}
    \Psi_{\Lambda}
    &=
    \left(\frac{-3}{128\eta u^{5}}\right)
        \left[ \frac{39}{2}\bar{\Lambda}
        u^{10} + \mathcal{O}(u^{12}) 
        \right]
    \,, \\
    \Psi_{\Xi}
    &=
    \left(\frac{-3}{128\eta u^{5}}\right)
    \left[ 
    \frac{75}{32}
    \bar{\Xi} 
    u^8 \log(u) 
    + 
    \mathcal{O}(u^{10}) 
    \right]
    \,.
\end{align}
\end{subequations}
All of this implies that our waveform is paramtrized by the following parameters
\begin{align}
    \Theta^a &= \left\{ \log(M), \log(\eta) , \bar{\Lambda}, \bar{\Xi}, t_c, \varphi_c, \log(\mathcal{A})\right\}\,.
\end{align}
We perform our Fisher analysis by using the sensitivity curves for advanced LIGO from~\cite{KAGRA:2013rdx}\footnote{\texttt{https://dcc.ligo.org/LIGO-T1800044/public}} and fix the SNR to $\rho = 100$, a value corresponding to a GW170817~\cite{LIGOScientific:2017vwq} like event detected at design sensitivity. 
We set $m_A=m_B=1.6\mathrm{M}_{\odot}$, $k_{2,A}=k_{2,B}=0.1$, $t_c=0$,
and $\varphi_c=0$; $\mathcal{A}$ is chosen to set $\rho=100$. 
We set $\left<\tau_A\right>=\left<\tau_B\right> \equiv \left<\tau\right> $.
\begin{figure}[h!]
    \centering
    \includegraphics[width = 0.85 \columnwidth]{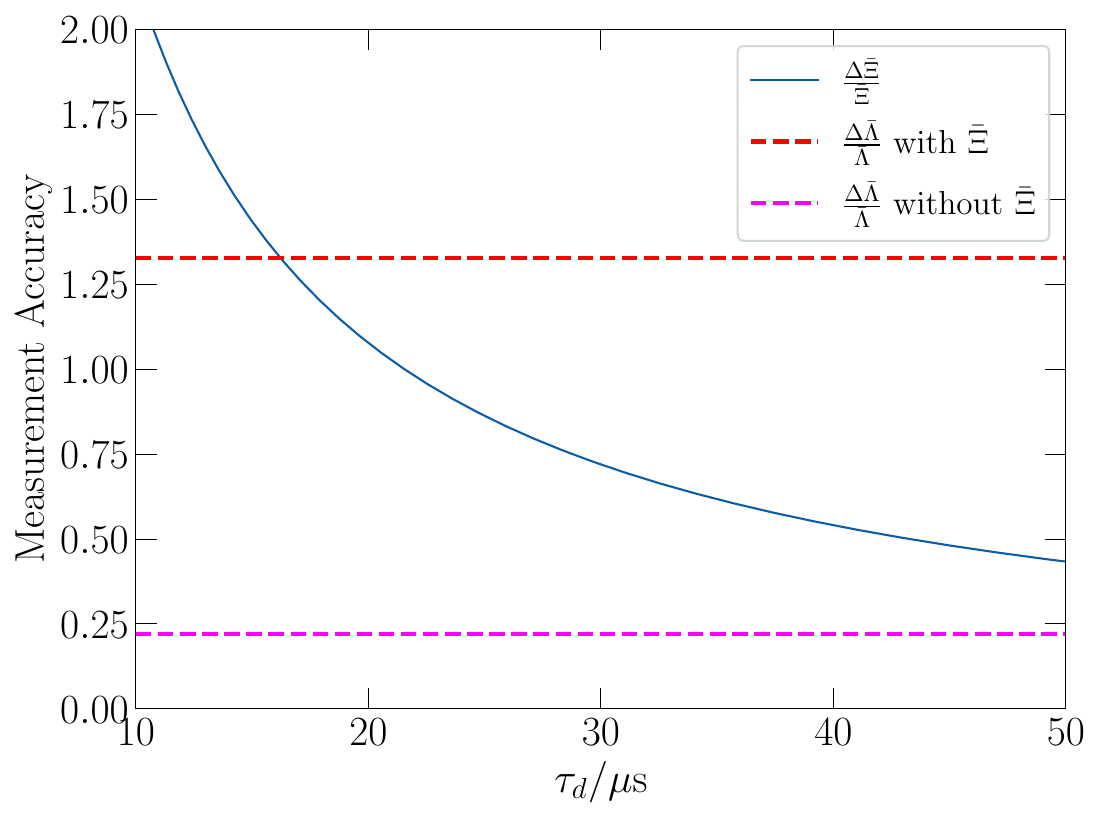}
    \caption{Measurement accuracy of the binary tidal deformabilities 
    for advanced LIGO with a fixed SNR of $100$ for an equal mass neutron star binary system. 
    The x-axis shows the injected values of $\tau$ and , where $\tau_A=\tau_B \equiv \tau$ is the tidal lag time (see Eq.~\eqref{eq:definition-tidal-deformabilities}).
    We see that the tidal lag deformability becomes measurable if the kinematic viscosity is greater than 
    $\tau_d \gtrsim 21\mu\mathrm{s}$.
    The presence of $\bar{\Xi}$ in the waveform introduces degeneracies, which increase the error on the measurement of 
    $\Delta \log \left( \bar{\Lambda} \right)$ by a factor of 6.}
    \label{fig:measurement-accuracy}
\end{figure}

The results of our Fisher analysis are presented in Fig.~\ref{fig:measurement-accuracy}, which shows the accuracy to which various tidal parameters can be measured as a function of the tidal lag parameter.
Since the effective adiabatic tidal deformability is independent of $\tau$, its measurability is independent of the injected value of $\tau$. 
However, the inclusion of the effective dissipative tidal deformability increases the dimensionality of the parameter space, therefore diluting the information content of the signal and deteriorating our ability to measure all parameters (including the adiabatic tidal ones). 
Perhaps more importantly, we see that if the injected tidal lag time is $\gtrsim 21 \mu \mathrm{s}$, then one may be able to measure the effective dissipative tidal deformability to better than $100\%$ precision. 
These measurement estimates of course scale inversely with the SNR, and therefore, a louder signal would be able to place a similarly stronger constraint, enabling measurements of the effective dissipative tidal deformability for even smaller values of the tidal lag parameter. 
This simple analysis reveals that there is a region in parameter space inside which one may be able to start to place meaningful constraints on $\Bar{\Xi}$, although these conclusions need to be confirmed with a full Bayesian analysis,
along with a self-consistent, relativistic calculation of $p_{2,A/B}$ with a realistic equation of state, to map the constraints on $\bar{\Xi}$ to constraints on the effective sources of viscosities.
For reference, using Eq.~\eqref{eq:tidal-lag-def}, a constraint on $\tau_d$ can be mapped onto a constraint $\stf{\nu}$.
An upper bound of $\tau_d \sim 20 \mu s$ would translate to an upper bound $\stf{\nu} \sim 3.7 \times 10^{16} \mathrm{cm^2/s}$ for the bulk viscous case ($p_2 \sim 0.05$) and an upper bound of $\stf{\nu} \sim 7.41 \times 10^{14} \mathrm{cm^2/s}$ for the shear viscous case ($p_2 \sim 5$).
The upper bound on the bulk viscous value approaches the limit set in Eq.~\eqref{eq:max-value-kinematic-viscosity} therefore, we expect that 3G detectors will be able to place more stringent constraints on bulk viscosity. 

\section{There are no out-of-equilibrium corrections to adiabatic deformations of stars
\label{sec:no-out-of-equlibrium-contribution-tidal-love}}

We have shown that two sets of tidal deformability coefficients,
$\Lambda_{A/B}$ and $\Xi_{A/B}$,
could have a roughly equal impact on the gravitational-wave phase for a neutron star binary, 
see Eqs.~\eqref{eq:Delta-Psi-Lambda1},~\eqref{eq:Delta-Psi-Lambda0}. 
We have also argued that to leading order,
$\Xi_{A/B}$ receives corrections due to the dissipative aspects of
the object's composition. 
In Sec.~\ref{sec:general-argument-no-viscous-corrections}, we provide a general argument that that there are no non-equilibrium/dissipative corrections to the
tidal deformability $\Lambda_{A/B}$, i.e.~that there are
no out-of-equilibrium contributions to the tidal Love number.
We additionally show that other adiabatic tidal quantities, like the
quadrupolar moment and moment of inertia of stars, 
are also unaffected by viscous corrections.

To strengthen our general argument, in Sec.~\ref{sec:no_viscous_corrections-quadrupole-intertia} and \ref{sec:no_viscous_corrections-Love} we explicitly show that the above results argument for two specific hyperbolic models of relativistic fluids--the BDNK \cite{Bemfica_2018,Bemfica:2020zjp,Kovtun_2019,Hoult:2020eho} and and Mueller-Israel-Stewart \cite{Bemfica:2019cop} fluid models--to supplement our general analysis (we provide a quick review of the two fluid models in 
Appendix~\ref{sec:review-relativistic-fluid-theories}).
We do this as in our general argument, we assume that there are no large gradients in the fluid solution. We focus on the BDNK and an Israel-Stewart model as those modes have been
shown to have a locally well-posed initial value problem.
While this in itself is a somewhat restrictive assumption,  we believe it is reasonable, and that our arguments could be extended to other fluid models as well, as we argue in Sec.~\ref{sec:extension_to_other_fluids}.

The upshot of this section is that we can view the two sets of tidal deformability
coefficients, $\Lambda_{A/B}$ and $\Xi_{A/B}$,
as parametrizing the leading equilibrium and out-of-equilibrium properties
of a neutron star, respectively.
Moreover, out-of-equilibrium effects do not affect the I-Love-Q relations,
which relate the relative values of the moment of inertia, Love number,
and quadrupole moment of neutron stars \cite{Yagi:2013bca,Yagi:2013awa,Yagi:2016bkt}.
We present our general argument for the absence of out-of-equilibrium corrections
to $\Lambda_{A/B}$ in Sec.~\ref{sec:general-argument-no-viscous-corrections}, 
and our more detailed argument in 
Secs.~\ref{sec:no_viscous_corrections-quadrupole-intertia}--\ref{sec:extension_to_other_fluids}.
Readers who are satisfied with our general argument may wish to skip those sections.

\subsection{General argument \label{sec:general-argument-no-viscous-corrections}}

Here we argue that on general, physical terms, there should be no viscous corrections
to tidal Love numbers, the quadrupolar moment, and the moment of inertia of a star.
These quantities are computed from time-independent solutions of the Einstein 
equations and the fluid equations of motion
\cite{1967ApJ...150.1005H,Hinderer:2007mb,Damour:2009vw,Binnington:2009bb}.
We expect that time-independent solutions to the fluid equations should be solutions
in thermal equilibrium, and we expect there to be no out-of-equilibrium contributions 
to the solution of the equilibrium Einstein equations.
As out-of-equilibrium effects are captured by the viscous/heat coefficients in a fluid
stress-energy tensor, we conclude that those coefficients should not contribute
to the solutions of the tidal Love number, quadrupole moment, and the moment
of inertia for a star.
In particular, the solutions to those quantities are determined entirely by the
perfect fluid part of the fluid stress-energy tensor.

We can slightly formalize this argument by showing that if the second law of
thermodynamics holds, the stress-energy tensor for stationary solutions should
reduce to that of a perfect fluid under a relatively weak set of assumptions
\cite{stewart1971non,Israel:1979wp,Becattini:2012tc,1976ApJ...208..873L,Becattini:2016stj}.
We parametrize the fluid stress-energy tensor as follows
\begin{align}
    T_{\mu\nu}
    =
    \mathcal{E}u_{\mu}u_{\nu}
    +
    \mathcal{P}\Delta_{\mu\nu}
    +
    \mathcal{Q}_{\mu}u_{\nu}
    +
    \mathcal{Q}_{\nu}u_{\mu}
    -
    2\eta\sigma_{\mu\nu}
    ,
\end{align}
where $u_{\mu}$ is the fluid four-vector,
$\Delta_{\mu\nu} \equiv g_{\mu\nu} + u_{\mu}u_{\nu}/c^2$ is the orthogonal
projection to $u_{\mu}$, $\mathcal{Q}_{\mu}$ is the fluid heat vector
(which is orthogonal to $u_{\mu}$), $\sigma_{\mu\nu}$ is the fluid
shear tensor (also orthogonal to $u_{\mu}$).
We interpret $\mathcal{E}$ as the generalized fluid energy density,
$\mathcal{P}$ as the generalized fluid  pressure, and $\eta$ as the shear viscosity coefficient.
For a perfect fluid, $\mathcal{Q}_{\mu}=0$, $\sigma_{\mu\nu}=0$,
$\mathcal{E}=e/c^2$, and $\mathcal{P}=p$, where $e$ is the internal
energy density and $p$ is the fluid pressure.
For more discussion on our notation, see Appendix~\ref{sec:review-relativistic-fluid-theories}.

The relativistic second law of thermodynamics is
\begin{align}
    \nabla_{\mu}S^{\mu}
    \geq 0
    ,
\end{align}
where $S^{\mu}$ is the fluid entropy current \cite{Israel:1976tn}.
We expect that time-independent solutions to the fluid conservation equations 
should be equilibrium solutions, so that $\nabla_{\mu}S^{\mu}=0$.
A general requirement for relativistic fluid flow is that the divergence of the
entropy current should satisfy \cite{Israel:1976tn,Bemfica:2020zjp} 
\begin{align}
\label{eq:second_law_thermo_gradients}
\nabla_{\mu}S^{\mu}
    =&
    \frac{\zeta}{T}\theta^2 
    +
    \frac{\eta}{T}\sigma_{\mu\nu}\sigma^{\mu\nu}
    +
    \frac{\kappa}{T^2}\mathcal{Q}_{\mu}\mathcal{Q}^{\mu} 
    \nonumber\\
    &
    +
    \mathcal{O}\left(\partial^3\right)
    \geq 0
    ,
\end{align}
where $\zeta$ is the bulk viscosity, 
$\kappa$ is the heat conductivity, $\theta$ is the fluid expansion, 
and $\mathcal{O}\left(\partial^3\right)$ refers to higher derivative terms,
which depend on the particular fluid model under consideration.
We note that 
$\sigma_{\mu\nu}\sigma^{\mu\nu}\geq0$, $\mathcal{Q}_{\mu}\mathcal{Q}^{\mu}\geq0$,
$\zeta\geq0$, $\eta\geq0$, and $\kappa\geq0$, so to leading order in derivatives,
$\nabla_{\mu}S^{\mu}\geq0$.

Fluid models can be viewed as long-wavelength effective field theories \cite{Romatschke:2017ejr}. 
From this point of view one
should demand that the second law holds \emph{to each order} in the gradient expansion,
and the breakdown of the second law at any given order would then signal a breakdown of the
fluid effective field theory.
Thus, assuming the fluid description of the given solution holds, 
we can demand that the following, stronger condition on the entropy current holds
\begin{align}
    \nabla_{\mu}S^{\mu}
    =&
    \frac{\zeta}{T}\theta^2 
    +
    \frac{\eta}{T}\sigma_{\mu\nu}\sigma^{\mu\nu}
    +
    \frac{\kappa}{T^2}\mathcal{Q}_{\mu}\mathcal{Q}^{\mu}\geq 0
    .
\end{align}
For equilibrium fluid solutions, one has that $\nabla_{\mu}S^{\mu}=0$,
and assuming that $\zeta>0$, $\eta>0$, and $\kappa>0$, 
this implies that $\theta=0$, $\sigma_{\mu\nu}=0$, 
and $\mathcal{Q}_{\mu}=0$, as $\theta^2$, $\sigma_{\mu\nu}\sigma^{\mu\nu}$,
and $\mathcal{Q}_{\mu}\mathcal{Q}^{\mu}$ are positive definite quantities.
Finally, if we \emph{assume} that $\mathcal{E}$ and $\mathcal{P}$ are respectively
equal to $e/c^2$ and $p$, plus terms that are proportional to $\theta$,
$\sigma_{\mu\nu}$, and time gradients of $e$ and $p$,
then the stress-energy tensor reduces to that of a perfect fluid for equilibrium solutions
(this holds for essentially all proposed viscous, relativistic fluid models;
for more discussion see Appendix~\ref{sec:review-relativistic-fluid-theories}).
This completes our argument. 

While this argument is physically plausible, discarding higher derivative
terms in the entropy relation of Eq.~\eqref{eq:second_law_thermo_gradients} 
is somewhat unsatisfactory. This is because while fluid models can
be interpreted as providing a long-wavelength, effective description of physical
systems in near-equilibrium, in practice it is common to find that steep
gradients form in solutions to fluid equations, as they form shock-like
solutions, where the widths of the shocks are regularized by the size of the
viscous/heat coefficients (for a review see for example \cite{Rezzolla-Book}).
Physically, we do not expect shock-like solutions to decrease the entropy, although it
has not yet been proven that any relativistic, hyperbolic extension of the Navier-Stokes
equations (that is, that incorporates viscous effects) satisfies this property.
Because of this, in the next subsections, we explicitly show that the higher-derivative corrections that appear in two different hyperbolic fluid models
do not contribute to the tidal Love number, the moment of inertia, and the quadrupole
moment (all equilibrium solutions) to the Einstein equations.

\subsection{The
quadrupole moment and the moment of inertia of (slowly rotating) neutron stars do not have viscous corrections
\label{sec:no_viscous_corrections-quadrupole-intertia}}

We first show there are no viscous corrections to the quadrupole moment
and moment of inertia of slowly-rotating neutron stars modeled as BDNK viscous fluids
(see Appendix ~\ref{sec:BDNK-Intro} for a quick review of BDNK fluids).
These are a class of hyperbolic, relativistic fluid models that consistently
include the effects of bulk viscosity, shear viscosity, and heat conduction.
We work in spherical polar coordinates; e.g. $x^{\mu}=\left(x^t,x^r,x^{\vartheta},x^{\varphi}\right)$.
In brief, we show that with the decomposition of 
Eqs.~\eqref{eq:circular_g} and \eqref{eq:circular_u}, the following holds:
\begin{enumerate}
\item The quantities $\theta$, 
    $\sigma_{tt}$, 
    $\sigma_{rr}$, 
    $\sigma_{t\varphi}$, 
    $\sigma_{r\vartheta}$, 
    $\sigma_{\vartheta\vartheta}$, 
    $\sigma_{\varphi\varphi}$, 
    $\mathcal{Q}_t$, and $\mathcal{Q}_{\varphi}$
    are all identically zero with a circular spacetime ansatz
    (that is, with Eq.~\eqref{eq:circular_g}, Eq.~\eqref{eq:circular_u},
    and all scalar quantities do not depend on $t$ or $\varphi$).
    Additionally $\mathcal{E}=e/c^2$ and $\mathcal{P}=p$, given that ansatz.
\item The quantities 
    $\sigma_{tr}$,
    $\sigma_{t\vartheta}$,
    $\sigma_{r\varphi}$,
    $\sigma_{\vartheta\varphi}$,
    $\mathcal{Q}_r$ and $\mathcal{Q}_{\vartheta}$
    are all zero because of the Einstein equations 
    $G_{\mu\nu} = (8\pi G/c^4)T_{\mu\nu}$, given the circular spacetime ansatz.
\end{enumerate}
From this, we conclude that the fluid stress-energy tensor reduces to that of a 
perfect fluid for solutions to the Einstein equations in circular spacetimes.
As a circular spacetime corresponds to the spacetimes used to compute the
quadrupole moment and moment of inertia for neutron stars \cite{1967ApJ...150.1005H}, 
we conclude that there are no viscous corrections to those quantities.
Below, we introduce the circular spacetime ansatz and 
describe the above calculations in more detail.

We start by showing that circular solutions to the 
BDNK equations of motion reduce to those of a perfect fluid. 
A circular spacetime is one that is stationary, axisymmetric and whose stress-energy tensor satisfies 
$t^{\mu}T^{[\nu}{}_{\alpha}t^{\alpha}\varphi^{\beta]}
=
\varphi^{\mu}T^{[\nu}{}_{\alpha}\varphi^{\alpha}t^{\beta]} =0$,
where $t^{\mu}$ and $\varphi^{\mu}$ are the timelike and spacelike
Killing vectors associated with stationarity and axisymmetry, respectively
(for reviews see \cite{Wald:1984rg,Gourgoulhon:2010ju}).
More intuitively, circular spacetimes are invariant under time translation, they
are axisymmetric, and they admit a spacetime metric that does not depend 
on the azimuthal direction $\varphi$.
Restricting ourselves to asymptotically flat spacetimes, 
we work in Hartle-Sharp coordinates in which the line element takes the form\cite{1967ApJ...147..317H,1967ApJ...150.1005H}
\begin{align}
\label{eq:circular_g}
    &
    g_{\alpha\beta}dx^{\alpha}dx^{\beta} 
    =
    -
    e^{-2\psi\left(r,\vartheta\right)}dt^2
    +
    e^{2\lambda\left(r,\vartheta\right)}dr^2
    \nonumber\\
    &
    +
    r^2A\left(r,\vartheta\right)^2
    \left[
        d\vartheta^2
        +
        \sin^2\vartheta\left( 
            d\varphi
            -
            \omega\left(r,\vartheta\right)dt
        \right)^2
    \right]
    .
\end{align}
In these coordinates, the two Killing vectors are $\partial_t$ and $\partial_{\varphi}$. 
We \emph{assume} the fluid velocity four-vector can be written as
\begin{align}
\label{eq:circular_u}
    u^{\mu}\partial_{\mu}
    =
    U\left(r,\vartheta\right)\left[ 
        \partial_t
        +
        \Omega\partial_{\varphi}
    \right] 
    \,,
\end{align}
where $\Omega$ is the angular velocity of the fluid, which we assume to be a constant.
For perfect fluids, one can show that the above equation must hold in circular spacetimes
due to the condition $t^{\mu}T^{[\nu}{}_{\alpha}t^{\alpha}\varphi^{\beta]}=0$
(e.g. \cite{Gourgoulhon:2010ju}). 
For BDNK fluids though, we have not been able to show that the fluid vector of
Eq.~\eqref{eq:circular_u} follows from the above condition,
since $u^{\mu}$ could potentially have nonzero $r$ or $\vartheta$ components, 
due to the presence of the heat vector $Q^{\mu}$ and the shear tensor. 
Instead, we view Eq.~\eqref{eq:circular_u} as
a reasonable, physical assumption for the form of the fluid velocity\footnote{If we assume the fluid is in thermal equilibrium, then
$\beta^{\mu}\equiv u^{\mu}/T$ must be a Killing vector 
\cite{stewart1971non,Israel:1979wp,Becattini:2012tc,Becattini:2016stj}. 
Said another way, in Eq.~\eqref{eq:circular_u} we assume that there can be no radial fluid flux that could be countered by the presence of a nonzero heat vector or shear tensor.
In this case, and assuming that the only Killing vectors are $\left(\partial_t\right)^{\mu}$ and
$\left(\partial_{\varphi}\right)^{\mu}$, 
we can conclude that $u^{\mu}$ takes the form of Eq.~\eqref{eq:circular_u}.}.
The metric of Eq.~\eqref{eq:circular_g} and
the fluid vector ansatz  of Eq.~\eqref{eq:circular_u} encompass, as special cases,
the metrics used for computing the moment of inertia and 
the quadrupole moment of static neutron stars \cite{stewart1971non}.
Finally, we assume all other fluid field quantities 
($\rho,\epsilon,p,\mu$ and $T$) are functions of $r$ and $\vartheta$ only.

With these assumptions, we can now prove that the expansion and the shear vanish. 
The computation of $\theta$ goes as follows 
\begin{align}
    \label{eq:theta_is_zero_circular}
    \theta 
    =&
    \frac{1}{\sqrt{-g}}\partial_{\mu}\left(\sqrt{-g}u^{\mu}\right)
    \nonumber\\
    =&
    \frac{1}{\sqrt{-g}}\partial_{t}\left(\sqrt{-g}u^{t}\right)
    +
    \frac{1}{\sqrt{-g}}\partial_{\varphi}\left(\sqrt{-g}u^{\varphi}\right)
    =
    0
    .
\end{align}
The second line follows from Eq.~\eqref{eq:circular_u}.
The last line follows from the fact that all field quantities are functions
only of $r,\vartheta$.
Given that $\theta=0$, the shear tensor simplifies to 
\begin{align}
    \sigma_{\mu\nu} 
    =
    \Delta^{\alpha}{}_{\mu}\Delta^{\beta}{}_{\nu}
    \left(
        \partial_{(\alpha}u_{\beta)}
        -
        \Gamma^{\gamma}_{\alpha\beta}u_{\gamma}
    \right)
    .
\end{align}
From the Einstein equations and the 
stress-energy tensor [Eq.~\eqref{eq:decomp_tmunu}], we have that
\begin{align}
    \Delta^{\alpha}{}_{\mu}\Delta^{\beta}{}_{\mu}G_{\alpha\beta}
    =
    \frac{8\pi G}{c^4}
    \Delta^{\alpha}{}_{\mu}\Delta^{\beta}{}_{\mu}T_{\alpha\beta}
    =
    -
    \frac{16\pi G}{c^4}
    \eta\sigma_{\mu\nu}
    .
\end{align}
The $(t,r)$, $(t,\vartheta)$, $(r,\varphi)$, and $(\vartheta,\varphi)$ components of 
$\Delta^{\alpha}{}_{\mu}\Delta^{\beta}{}_{\nu}G_{\alpha\beta}$ are zero,
which implies from the Einstein equations that 
$\sigma_{tr}$, $\sigma_{t\vartheta}$, $\sigma_{r\varphi}$, and
$\sigma_{\vartheta\varphi}$ are all zero.
The components $\partial_{(t}u_{t)}$, $\partial_{(t}u_{\varphi)}$,
and $\partial_{(\varphi}u_{\varphi)}$
are zero as the components of $u_{\mu}$ only depend on $r$ and $\vartheta$.
Since only $u_t$ and $u_{\varphi}$ are nonzero, the following partial derivatives are also zero:
$\partial_{(r}u_{r)}$, $\partial_{(r}u_{\vartheta)}$, and 
$\partial_{(\vartheta}u_{\vartheta)}$. From the form of the metric
Eq.~\eqref{eq:circular_g}, and fluid four-velocity, Eq.~\eqref{eq:circular_u},  
we find that
$\Gamma^{\gamma}_{tt}u_{\gamma}$,
$\Gamma^{\gamma}_{t\varphi}u_{\gamma}$,
$\Gamma^{\gamma}_{\varphi\varphi}u_{\gamma}$,
$\Gamma^{\gamma}_{rr}u_{\gamma}$,
$\Gamma^{\gamma}_{r\vartheta}u_{\gamma}$, and
$\Gamma^{\gamma}_{\vartheta\vartheta}u_{\gamma}$ are all zero.
This implies that the remaining components of the shear are zero.
Putting everything together, we have 
\begin{align}
    \sigma_{\mu\nu}
    =
    0
    .
\end{align}

Lastly, we turn to the heat vector. 
From the Einstein equations and the 
stress-energy tensor of Eq.~\eqref{eq:decomp_tmunu}, we have that
\begin{align}
    \label{eq:u_projection_tmunu} 
    u^{\nu}G^{\mu}{}_{\nu}
    =
    \frac{8\pi G}{c^4} 
    u^{\nu}T^{\mu}{}_{\nu} 
    =
    -
    \frac{8\pi G}{c^4} 
    \left( 
        \mathcal{E}u^{\mu}
        +
        \mathcal{Q}^{\mu}
    \right) 
    .
\end{align}
The Einstein tensor components $u^{\mu}G^r_{\mu}$ and $u^{\mu}G^{\vartheta}_{\mu}$
are zero for the metric of Eq.~\eqref{eq:circular_g}.
From the Einstein equations and $u^r=0$ and $u^{\vartheta}=0$ (from Eq.~\eqref{eq:circular_u}), 
we see that $\mathcal{Q}^r=0$ and $\mathcal{Q}^{\vartheta}=0$,
which imply that $\mathcal{Q}_r=0$ and $\mathcal{Q}_{\vartheta}=0$.
To compute $\mathcal{Q}_t$ and $\mathcal{Q}_{\varphi}$, we consider
the individual components of $\mathcal{Q}_{\mu}$
(see Eq.~\eqref{eq:Q-def-general-param}).
The only nonzero components of
$\Delta^{\mu}{}_{\nu}$ that contain a $t$ and/or a $\varphi$ index are
$\Delta^t{}_t$, $\Delta^t{}_{\varphi}$, $\Delta^{\varphi}{}_t$, and $\Delta^{\varphi}{}_{\varphi}$:
\begin{subequations}
\begin{align}
    \Delta^t{}_t
    =&
    1
    -
    \left(
        e^{-2\psi}
        +
        r^2A^2 \left(\Omega-\omega\right)\omega\sin^2\vartheta
    \right)
    U^2
    \\
    \Delta^t{}_{\varphi}
    =&
    \left(\Omega-\omega\right)A^2U^2r^2\sin^2\vartheta
    \\
    \Delta^{\varphi}{}_t
    =&
    -
    \left(
        e^{-2\psi}
        +
        A^2\left(\Omega-\omega\right)\omega r^2\sin^2\vartheta
    \right)
    \Omega U^2
    \\
    \Delta^{\varphi}{}_{\varphi}
    =&
    1
    +
    \left(\Omega-\omega\right)\Omega A^2U^2 r^2\sin^2\vartheta
    .
\end{align}
\end{subequations}
For any function $f\left(r,\vartheta\right)$, we have 
\begin{align}
    \Delta^{\mu}{}_t\nabla_{\mu}f
    =
    0
    ,\qquad
    \Delta^{\mu}{}_{\varphi}\nabla_{\mu}f
    =
    0
    ,
\end{align}
Since $p$ and $\mu/T$ are functions only of $r$ and $\vartheta$,
we conclude that their projected derivatives do not contribute to the 
$t$ and $\varphi$ components of $\mathcal{Q}_{\mu}$.
One can also show that $u^{\mu}\nabla_{\mu}u_{t}=0$ and
$u^{\mu}\nabla_{\mu}u_{\varphi}=0$.
From the definition for $\mathcal{Q}_{\mu}$ (Eq.~\eqref{eq:Q-def-general-param}),
we conclude that $\mathcal{Q}_t=0$ and $\mathcal{Q}_{\varphi}=0$.
Putting everything together, we have that
\begin{align}
    \mathcal{Q}_{\mu} = 0
    .
\end{align}

From $\vartheta=0$, $\sigma_{\mu\nu}=0$, and $\mathcal{Q}_{\mu}=0$, 
we conclude that the on-shell stress-energy tensor 
reduces to that of a perfect fluid
\begin{align}
\label{eq:perfect_fluid_se_tensor}
    T_{\mu\nu}
    =
    \frac{e}{c^2} u_{\mu}u_{\nu}
    +
    p\Delta_{\mu\nu}
    .
\end{align}
The metric and fluid vectors of Eqs.~\eqref{eq:circular_g} and 
\eqref{eq:circular_u} take the standard form used
in computations of the moment of inertia and quadrupole
moment of neutron stars \cite{1967ApJ...150.1005H}. 
Since the stress-energy tensor reduces to that of a perfect fluid for this spacetime,
one can see that, by following the same steps as in \cite{1967ApJ...150.1005H}
to derive the moment of inertia and quadrupole, there are no viscous corrections
to those two quantities. 
Finally, we also note a special case of the above results:
there are no viscous corrections to static fluid solutions.
Namely, there are no viscous corrections to the 
Tolman-Oppenheimer-Volkoff equations for static stars \cite{PhysRev.55.374}.

\subsection{The
(adiabatic) tidal Love number of static neutron stars do not have viscous corrections
\label{sec:no_viscous_corrections-Love}}

We next compute the (adiabatic) tidal Love numbers for static neutron stars modeled as BDNK fluids
(see Appendix ~\ref{sec:BDNK-Intro} for a quick review of BDNK fluids).
As we discussed in Sec.~\ref{sec:effective-point-particle},
the tidal Love numbers describe the linear tidal response of a star
when perturbed by a stationary external gravitational field
\cite{Hinderer:2007mb,Damour:2009vw,Binnington:2009bb}.
As we discussed in Sec.~\ref{sec:viscosity-PN-order-phase},
in a neutron star binary, each star will be perturbed by the gravitational field
produced by its companion, which will deform the star. While the gravitational field
will slowly change in the rest frame of the star, the timescale of that change is
slow compared to the internal dynamical timescale of each star, so the leading-order contribution to the stars' tidal response is the adiabatic tidal Love number, 
provided the star is not resonantly excited by its
companions gravitational field (c.f. Eq.~\eqref{eq:Q_freq_sho}).
Unless otherwise noted, 
in this section we refer to adiabatic tidal Love numbers as just Love numbers for brevity.

For static neutron stars, the Love numbers are classified by how the
imposed gravitational field transforms under parity \cite{Damour:2009vw,Binnington:2009bb}.
The axial Love numbers describe the linear response to 
an odd-parity, external gravitational field, while the polar Love numbers describe the linear
response of the star to an even-parity external gravitational field. 
While only the even-parity quadrupolar Love number is large
enough to observably affect the dynamics of a neutron star binary 
\cite{Damour:2009vw,Binnington:2009bb} with current detectors,
we consider both the axial and polar Love numbers here.

We denote background quantities with a $(0)$ subscript (e.g. $g_{(0)}^{\mu\nu}$
is the background metric tensor),
use $\delta$ to denote
a linear perturbation (e.g. $\delta g^{\mu\nu}$ is the linear perturbation of the metric),
and write the combinations of the background and linear fields with no subscript/change
(e.g. $g^{\mu\nu}=g_{(0)}^{\mu\nu} + \delta g^{\mu\nu}$).
Following earlier work, 
we use the Regge-Wheeler gauge for the linearized metric perturbation 
\cite{Hinderer:2007mb,Damour:2009vw,Binnington:2009bb}.
Our notation for the spherical harmonics follows \cite{Martel:2005ir},
except that we put a bracket around the vector/tensor spherical harmonics to
distinguish the angular numbers $(\ell,m)$ from the tensorial indices.
We take the background metric and fluid velocity to be
given by Eqs.~\eqref{eq:circular_g} and~\eqref{eq:circular_u}, with
$A=1$, $\omega=0$, $\Omega=0$, and $\psi$, $\lambda$, and $U$ are only functions of $r$:
\begin{subequations}
\begin{align}
    \left(g_{(0)}\right)_{\mu\nu}dx^{\mu}dx^{\nu}
    =&
    -
    e^{-2\psi\left(r\right)}dt^2
    +
    e^{2\lambda\left(r\right)}dr^2
    \nonumber\\
    &
    +
    r^2\left(d\vartheta^2+\sin^2\vartheta d\varphi^2\right)
    ,\\
    \left(u_{(0)}\right)^{\mu}\partial_{\mu}
    =&
    e^{\psi\left(r\right)}\partial_t
    .
\end{align}
\end{subequations}
This is a special case of a circular spacetime, so using the
results in Sec.~\eqref{sec:no_viscous_corrections-quadrupole-intertia}, 
we conclude that there are no viscous corrections to spacetimes of this form.
\subsubsection{Axial Love numbers\label{sec:axial_love_numbers}}
We first consider the axial Love numbers.
For axial perturbations, the metric perturbation takes the form \cite{Binnington:2009bb}\footnote{Our results still hold if we let $\delta u^{\alpha}$ have a small axial component, 
which is taken to zero for the calculation for the
axial Love number \cite{Landry:2015cva,Pani:2018inf}. 
This is because for $m=0$ perturbations, $\delta u^{\alpha}$
only gains a nonzero $\varphi$ component, and the equations still reduce to the
circular ansatz Eq.~\eqref{eq:circular_u}.}.
\begin{subequations}
\label{eq:axial_pert}
\begin{align}
    \delta g_{\alpha\beta}dx^{\alpha}dx^{\beta}
    =&
    -
    2h_0(r)\left[X^m_{\ell}\left(\vartheta,\varphi\right)\right]_Bdtdx^B
    ,\\
    \delta u^{\alpha}\partial_{\alpha}
    =&
    0
    ,
\end{align}
\end{subequations}
where $[X^m_{\ell}]_B$ are axial vector spherical harmonics.
Axial perturbations for all scalar quantities, such as the rest-mass
energy density $\rho$, are zero.

We can obtain the expressions for 
$g_{\mu\nu}$ and $T_{\mu\nu}$ when the angular number $m\neq0$  
by rotating (along the azimuthal direction) their solutions evaluated at $m=0$.
(c.f. \cite{1967ApJ...149..591T}).  
Setting $m=0$, we see that the axial spacetime of Eq.~\eqref{eq:axial_pert}
represents a special case of the circular metric of Eq.~\eqref{eq:circular_g} and
the circular fluid four-velocity of  Eq.~\eqref{eq:circular_u}, as the
$\varphi$ dependence drops out of those quantities.
We can then make use of our results on circular spacetimes to conclude that
$\theta=0$, $\sigma_{\mu\nu}=0$, and $\mathcal{Q}_{\mu}=0$.
From this, we see that the perturbed 
stress-energy tensor reduces to that of a perfect fluid. 
We conclude that are no viscous corrections to the axial Love numbers: they
are purely determined by the perfect fluid components of the stress-energy tensor.

\subsubsection{Polar Love numbers\label{sec:polar_love_numbers}}
We next consider the polar Love numbers. 
The metric and fluid velocity for polar perturbations take the form
\cite{Hinderer:2007mb,Damour:2009vw,Binnington:2009bb}
\begin{subequations}
\label{eq:polar_pert}
\begin{align}
    &
    \delta g_{\alpha\beta}dx^{\alpha}dx^{\beta} 
    =
    -
    \Big[
        e^{-2\psi(r)} H_0(r)dt^2 
        \nonumber\\
        &
        \qquad 
        +
        2 H_1(r) dtdr 
        +
        e^{2\lambda(r)}H_2(r) dr^2
        \nonumber\\
        &
        \qquad 
        +
        r^2K(r)\Omega_{AB}d\vartheta^Ad\vartheta^B
    \Big]
    Y^m_{\ell}\left(\vartheta,\varphi\right)
    ,\\
    &
    \delta u^{\alpha}\partial_{\alpha}
    =
    -
    \frac{1}{2}e^{\psi(r)}H_0(r)Y^m_{\ell}\left(\vartheta,\varphi\right)
    \partial_t 
    .
\end{align}
\end{subequations}
where $Y^m_{\ell}$ is the scalar
spherical harmonic and $\Omega_{AB}$ is the metric of the two-sphere. 
In addition, all scalar fluid quantities, such as $\rho$, are linearly perturbed. For example,
\begin{align}
    \delta\rho 
    =
    \rho_1\left(r\right)Y^m_{\ell}\left(\vartheta,\varphi\right)
    .
\end{align}

Unlike the case for axial perturbations,
even when $m=0$ the metric has a nonzero $(t,r)$ component, so it
does not reduce to a circular metric.
Instead, we consider a more direct approach to show that the
perturbed stress-energy tensor reduces to that of a perfect fluid.
As our derivation is more involved than it is for the axial Love numbers,
we first outline its main steps:
\begin{enumerate}
    \item First, we show that the equations of motion for $H_{0}$, $H_{2}$, and $K$ 
    remain unchanged from those of the perfect fluid case. 
    \item Next, we show that $H_1$ must be zero outside the star, it does not
    couple to any of the other metric variable, and it does not affect the equations
    of motion for $e$, $p$, $u^{\mu}$, or the metric variables $H_0$, $H_2$, and $K$.
    \item These results imply that the metric remains the same as in the 
    perfect fluid case outside the star, and the master equation for
    $\delta g_{tt}$ is the same as for the perfect fluid case.
    As the tidal Love numbers are defined through the multiple expansion
    of $\delta g_{tt}$ in a buffer zone far from the stellar surface and from 
    the source of the external field, we conclude the tidal Love numbers remain
    unaffected by viscosity.
\end{enumerate}

We now present each of these steps in more detail.
By inserting the values for $g_{\mu\nu}$ and $u^{\mu}$ 
from Eq.~\eqref{eq:polar_pert} and Eq.~\eqref{eq:axial_pert} into the
expressions for $\theta$ and $\sigma_{\mu\nu}$, 
one can show through direct calculation that $\theta=0$, $\sigma_{\mu\nu}=0$,
$\mathcal{Q}_t=0$, and $\mathcal{Q}_{\varphi}=0$.
While $\theta=0$ and $\sigma_{\mu\nu}$ for polar perturbations,
the heat vector components $\mathcal{Q}_r$ and $\mathcal{Q}_{\vartheta}$
generally remain nonzero (when $m=0$, which, without losing generality,
we keep to simplify our discussion).  
Despite this, we can follow the steps of \cite{Hinderer:2007mb} 
to derive a master equation for the metric perturbation $H_0$. 
Moreover, the master equation is unaffected by the nonzero components of the perturbed heat vector.
To show this, we notice that, as all metric components do not depend on
time, $\theta=0$, and only the $t$ component of $\delta u^{\mu}$ is nonzero,
we have that (see Eq.~\eqref{eq:E-def-general-param} and Eq.~\eqref{eq:P-def-general-param})
\begin{subequations}
\begin{align}
    \delta\mathcal{E}
    =&
    \frac{\delta e}{c^2}
    ,\\
    \delta\mathcal{P}
    =&
    \delta p
    .
\end{align}
\end{subequations}
The only nonzero component of $\delta\Delta^{\mu}{}_{\nu}$ is
\begin{align}
    \delta\Delta^t{}_r
    =
    -
    e^{2\psi}H_1
    .
\end{align}
Using that $\theta=0$, $\delta\sigma^{\mu}{}_{\nu}=0$, and $\mathcal{Q}_{(0)\mu}=0$, 
the perturbed stress-energy tensor is then 
\begin{align}
    \delta T^{\mu}{}_{\nu}
    =&
    \frac{\delta e}{c^2}u_{(0)}^{\mu}u_{(0)\nu}
    +
    \frac{e_{(0)}}{c^2}\left(\delta u^{\mu}u_{(0)\nu} + u_{(0)}^{\mu}\delta u_{\nu}\right)
    \nonumber\\
    &
    +
    \delta p\left(\Delta_{(0)}\right)^{\mu}{}_{\nu}
    +
    p_{(0)}\delta \Delta^{\mu}{}_{\nu}
    +
    u_{(0)}^{\mu}\delta\mathcal{Q}_{\nu}
    .
\end{align}
Putting everything together, we see that
\begin{subequations}
\begin{align}
    \delta T^t{}_t
    =& 
    -
    \frac{\delta e}{c^2}
    ,\\
    \delta T^r{}_r
    =
    \delta T^{\vartheta}{}_{\vartheta}
    =
    \delta T^{\varphi}{}_{\varphi}
    =&
    \delta p
    ,\\
    \delta T^r{}_{\vartheta}
    =&
    0
    .
\end{align}
\end{subequations}
The main result to notice is that these perturbed metric components take
the same form as they do for a perfect fluid: the heat vector does not enter the equations,
nor do any other BDNK corrections.
We then notice that even though $H_1$ is generally nonzero, 
it does not enter $\delta G^{t}{}_{t}$, 
$\delta G^{r}{}_{r}$, $\delta G^{\vartheta}{}_{\vartheta}$, $\delta G^{\varphi}{}_{\varphi}$,
or $\delta G^{r}{}_{\vartheta}$.
Ultimately, we find that only the $(t,t)$, $(r,r)$, $(\vartheta,\vartheta)$, and $(\varphi,\varphi)$
components of the Einstein equations enter the calculation of the polar
tidal Love number.
From this we conclude that quantity receives no viscous corrections.
We provide more details of our argument in Appendix \ref{sec:stationary-polar-perturbation}.

\subsection{Extension of our results to other viscous fluid models \label{sec:extension_to_other_fluids}}

Looking back at Sec.~\ref{sec:no_viscous_corrections-quadrupole-intertia} and 
Sec.~\ref{sec:no_viscous_corrections-Love}, we see that our argument that the quadrupole moment, the
moment of inertia, and the adiabatic Love numbers are not affected by viscosity in BDNK theories
rests on the following observations about the computation of those quantities:
\begin{enumerate}
    \item The fluid expansion $\theta$ is identically zero.
    \item The shear tensor contains some components that are identically zero.
        For the components that are not identically zero, they can be shown to be zero
        from the Einstein equations
        $\Delta_{\mu}{}^{\alpha}\Delta_{\nu}{}^{\beta}G_{\alpha\beta}
        =-\left(16\pi G/c^4\right)\eta \sigma_{\mu\nu}$,
        and from the fact that the Einstein tensor is zero for those components. 
    \item The components $\mathcal{Q}^t$ and $\mathcal{Q}^{\varphi}$ of the heat vector 
        can be shown to be identically zero.
    \item The generalized fluid energy and pressure reduce to the energy density and pressure of 
    a perfect fluid when $\theta=0$ and the fluid flow and spacetime are time independent.
\end{enumerate}
The first two requirements follow from kinematical arguments and from the form of the
Einstein equations.
Only the latter two requirements depend on the specific nature of the fluid stress-energy
tensor and the equations of motion.

We end this section by showing that these conditions hold for an Israel-Stewart model that has been 
shown to be causal and strongly hyperbolic (for a review of this model, see Appendix~\ref{sec:MIS-intro}).
For circular spacetimes $u^{\alpha}\nabla_{\alpha}\Pi=0$, while for polar perturbations
$u^{\alpha}\nabla_{\alpha}\Pi=0$.
In both cases, $\theta=0$ as well, so Eq.~\eqref{eq:relaxation_is_bulk} reduces to 
\begin{align}
    \Pi + \lambda\Pi^2 = 0 
    .
\end{align}
The smaller-in-magnitude solution to this is $\Pi=0$, which implies that the stress-energy
tensor reduces to that of a perfect fluid. We conclude that there are no viscous corrections
to the quadrupole moment, moment of inertia, or the adiabatic tidal Love numbers.

\section{Conclusions}\label{sec:conclusions}

In this article we have addressed the impact of tidal dissipation on the dynamics of neutron star binaries.
We have shown that the dissipative tidal deformability enters at 4PN order in the gravitational-wave phase,
one full PN order lower than equilibrium effects enter through adiabatic tidal deformability. 
Moreover, as for the conservative tidal deformability, the dissipative tidal deformability
receives a large finite size correction, which makes it potentially measurable or constrainable with
current ground-based gravitational-wave detectors.
We additionally showed that there are no out-of-equilibrium contributions to the
tidal Love number, the quadrupole moment, and the moment of inertia of a neutron star.
This implies that there are no out-of-equilibrium contributions to the
I-Love-Q relations \cite{Yagi:2013bca,Yagi:2013awa,Yagi:2016bkt}, 
and that the measurement of the two tidal deformability
parameters (the equilibrium one and the new viscous one) probe \textit{distinct} physical processes inside a neutron star.

This preliminary study opens up several avenues for future work.
First, we have not provided a complete derivation of the numerical values that $\Xi_A$ could take for a star. 
This expression for $\Xi_A$ (Eq.~\eqref{eq:definition-tidal-deformabilities}) contains an unknown parameter $p_{2}$, 
the value of which needs to be computed for realistic nuclear EOS to map a constraint on $\bar{\Xi}$ to a constraint on bulk or shear viscosity.

We have only provided a preliminary estimate of the effects of $\Xi_A$ on the gravitational-wave phase. 
Determining how easily the size of out-of-equilibrium, dissipative effects could be constrained/measured
with current and future gravitational-wave data will require a more detailed analysis.
A more thorough investigation of systematic effects in the gravitational-wave phase is required to address the robustness of the measurability of $\Xi_A$.

Continuing with potential systematic effects in measuring $\Xi_A$,
we have argued that higher-order derivative corrections in the relation between the quadrupolar
neutron-star response and the imposed gravitational field, Eq.~\eqref{eq:general_time_expansion}
are much smaller than the leading two terms.
While subleading terms may be numerically smaller, they could still systematically bias measurements
of $\Lambda_A$ and $\Xi_A$ if not taken into account.
This has already been argued to be the case for $\Lambda_A$ \cite{Pratten:2021pro} for
higher-order even-derivative contributions; it would
then be interesting to consider higher-order odd-derivative corrections in the
expansion Eq.~\eqref{eq:general_time_expansion}, and determine their effect on
measurement of $\Xi_A$.

Even if the effective kinematic viscosity of neutron star matter is too small
to be measured with gravitational waves, determining an upper bound on that quantity 
(through an upper bound on $\Xi_A$) may still lead to new astrophysically relevant
insights on the properties of neutron stars. 
For example, an upper bound on the kinematic viscosity will place a lower bound on
the damping time of the $f-$mode stellar oscillations, which play a leading role
in the dynamical tides of neutron stars \cite{Steinhoff:2016rfi,Hinderer:2016eia}.
An upper bound on the kinematic viscosity will also inform estimates of the damping time
of other oscillation patterns of neutron stars
(e.g. the $p$ and $g-$modes \cite{1987ApJ...314..234C,1990ApJ...363..603C}). 
Finally, we note that there is a wide range of 
estimates for the relative importance of out-of-equilibrium effects during
the inspiral and merger of neutron stars 
\cite{1992ApJ...398..234K,Jones:2001ya,Shternin:2008es,Arras:2018fxj,Most:2021zvc}.
Given this, even a modest gravitational-wave constraint on $\Xi_A$ 
may inform work on the non-equilibrium nuclear physics of neutron stars.

\begin{acknowledgements}
JR, AH, and NY acknowledge support from the Simons Foundation through Award number 896696 and support from NSF Award PHY-2207650.
We thank Rohit Chandramouli, Mark Alford, \'{E}anna Flanagan,  Geraint Pratten, and Patricia Schmidt for discussions
and correspondence about this article.
We thank Eric Poisson and Clifford Will for helpful correspondence on
Newtonian calculations of the dissipative tidal deformability.
We additionally thank the anonymous referee for their detailed and helpful comments on this article.
\end{acknowledgements}
\appendix
\input{appendix-binary-only.tex}
\bibliography{ref}
\end{document}

%% file: appendix-binary-only.tex
\section{Derivation of the gravitational phase formula}\label{appendix:derivation-phasing}
In this appendix, we provide a derivation of Eq.~\eqref{eq:d2Psi-domega2} starting from Eq.~\eqref{eq:d2Psi-domega2-original}.
Let us first rewrite Eq.~\eqref{eq:d2Psi-domega2-original} as 
\begin{align}\label{eq:Psi-f}
    \Psi(f) &= 2 \pi f t(f) - 2 \phi(f) \,,
\end{align} 
where
\begin{align}
    t(f) &\equiv t_c + \int^{f/2} \frac{1}{\dot{F}} dF \,, \nonumber \\
    \label{eq:tf-def}
    &= t_c + \int^{f/2}  \frac{E_{tot}'(F)}{\dot{E}_{tot}}  dF\,,
\end{align}
and
\begin{align}
    \phi(f)&\equiv \phi_c + 2 \pi \int^{f/2} \frac{F}{\dot{F}} dF \,,\nonumber \\
    &= \phi_c + 2 \pi \int^{f/2}  \frac{E_{tot}'(F)}{\dot{E}_{tot}} F dF\,.
\end{align}
Taking the $f$ derivative of Eq.~\eqref{eq:Psi-f} we get
\begin{align}
    \Psi'(f) &= 2 \pi t(f) + 2 \pi f t'(f) - 2 \phi'(f) 
    = 
    2 \pi t(f)\,,
\end{align}
where the last two terms of the first equality cancel because $2 \pi f (dt/df) = 2 (d\phi/dt) (dt/df) = 2 \phi'$. Taking the second derivative and using Eq.~\eqref{eq:tf-def}, we find Eq.~\eqref{eq:d2Psi-domega2}, namely
\begin{align}
    \frac{d^2 \Psi}{df^2} &= 2 \pi t'(f) = \frac{\pi}{\dot{E}_{tot}} \left.\frac{dE_{tot}}{dF}\right|_{F=\frac{f}{2}} =
    \frac{2\pi}{\dot{E}_{tot}}\frac{dE_{tot}}{df}
    \,.
\end{align}

\section{A brief review of hyperbolic relativistic fluid theories
\label{sec:review-relativistic-fluid-theories}}

A challenge to constructing relativistic, viscous fluid models is that
relativistic theories must be \emph{causal}--the speed of all fluid modes should be
bounded by the speed of light. Moreover (as with the Newtonian Navier-Stokes equations)
the equations of motion for the model should have a well-posed initial value problem,
which for causal theories implies the equations must be strongly hyperbolic \cite{Sarbach:2012pr}.
The earliest proposals for a relativistic generalizations of the Navier-Stokes equations
were done by Eckart \cite{PhysRev.58.919} and Landau and Lifshitz \cite{1959flme.book.....L}. 
The exact equations of motion for these models are neither causal,
nor do they admit stable equilibrium states \cite{1985PhRvD..31..725H,Pu:2009fj},
so they are not seen as physically viable theories of viscous fluids. 

An alternative approach to constructing relativistic, viscous fluid models
was pioneered by M\"{u}ller (in the non-relativistic setting), Israel, and Stewart 
(in the relativistic setting)
\cite{1967ZPhy..198..329M,Israel:1976tn,ISRAEL1976213,Israel:1979wp}. 
These \emph{extended (thermodynamic)} models promote the fluid viscosity and heat 
conduction terms to independent fields, which relax to their physical values via auxiliary 
equations of motion (for a review, see for example \cite{Rezzolla-Book}). 
The linearization of Israel-Stewart models about equilibrium states
have been shown to be causal and stable \cite{1983AnPhy.151..466H}. 
In the heavy-ion literature, popular extensions of the original Israel-Stewart models
include the DNMR model \cite{Denicol_2012} and the (r)BRSSS model \cite{Baier:2007ix}.
Stability and hyperbolicity results for these latter two theories have been shown
to hold for special backgrounds (e.g. about thermal equilibrium), but there are not yet
any general proofs of well-posedness for general solutions. 

In the first-order approach, the fluid stress-energy tensor is expanded to first order
in gradients about the perfect fluid model. 
From the effective field theory point of view \cite{Romatschke:2017ejr}, 
first order models can be thought of as parameterizing
the leading order non-equilibrium corrections to the fluid stress-energy tensor,
in terms of a gradient expansion about a perfect fluid background.
Bemfica, Disconzi, Noronha, and Kovtun first observed that  
\cite{Kovtun_2019,Bemfica:2020zjp} within the class of all first-order fluid models,
there is a subclass of models such that the equations of motion are
strongly hyperbolic, and modally stable about thermal equilibrium.
This subclass of models are called BDNK fluids.

\subsection{First order relativistic, viscous hydrodynamics\label{sec:BDNK-Intro}}

We first review the BDNK fluid model, as presently it is the only relativistic, viscous
fluid model that incorporates the bulk viscosity, shear viscosity, and heat conduction
fluid coefficients, and has been shown to be causal and have a locally well-posed initial
value problem, even for non-equilibrium fluid flows.

We first decompose the stress-energy tensor $T_{\mu\nu}$
into components parallel and perpendicular to the fluid four-velocity $u^{\mu}$:
\begin{align}
    \label{eq:decomp_tmunu}
    T_{\mu\nu} 
    &= 
    \mathcal{E} u_{\mu} u_{\nu}
    +
    \mathcal{P} \Delta_{\mu \nu}
    +
    2\mathcal{Q}_{(\mu}u_{\nu)}
    +
    \mathcal{T}_{\mu\nu}
    \,.
\end{align}
where $\Delta_{\mu \nu} \equiv g_{\mu\nu} + c^{-2} u_{\mu} u_{\nu}$ is the projection tensor.
We also have $u_{\mu}\mathcal{Q}^{\mu}=0$, $u_{\mu}\mathcal{T}^{\mu\nu}=0$, and
$g_{\mu\nu}\mathcal{T}^{\mu\nu}=0$.
For a relativistic perfect fluid, $\mathcal{E}=e$, where $e$ is the total energy density, 
$\mathcal{P}=p$, where $p$ is the pressure, and $\mathcal{Q}^{\mu}=0$, $\mathcal{T}_{\mu\nu}=0$.
In the BDNK approach, one expands the quantities $\mathcal{E}$, $\mathcal{P}$, $\mathcal{Q}_{\mu}$,
and $\mathcal{T}_{\mu\nu}$ in a gradient expansion about the equilibrium fluid quantities
$e$, $p$, and the fugacity $\mu/T$ (here $\mu$ is the chemical potential and $T$ is the temperature).
Following BDN \cite{Bemfica:2020xym}, who proved the strong hyperbolicity of the model, 
we restrict the fluid current to the Eckart form 
\begin{align}
    J^{\mu}
    =
    \rho u^{\mu}
    ,
\end{align}
that is, we do not consider a gradient expansion of the fluid four-current.
Given this, the most general, non-redundant expansion of the fluid variables is
(our presentation follows \cite{HegadeKR:2023glb})
\begin{subequations}\label{eq:parameterization-general}
\begin{align} 
\label{eq:E-def-general-param}
    \mathcal{E} 
    \equiv&
    \frac{e}{c^2} + \frac{\tau_{\epsilon,1}}{c^2} u^{\alpha} \nabla_{\alpha} \epsilon 
    +
    \frac{\tau_{\epsilon,2}}{c^2} \theta
    \nonumber \\
    &+
    \frac{\tau_{\epsilon,3}}{c^4} u^{\alpha}\nabla_{\alpha} \left(\frac{\mu}{T}\right)\,, \\
\label{eq:P-def-general-param}
    \mathcal{P} 
    \equiv& 
    p 
    -
    \zeta\theta
    \nonumber\\
    &
    +
    \tau_{p,1} u^{\alpha} \nabla_{\alpha} \epsilon
    +
    \tau_{p,2} \theta
    +
    \frac{\tau_{p,3}}{c^2} u^{\alpha}\nabla_{\alpha} \left(\frac{\mu}{T}\right) \,,\\
\label{eq:Q-def-general-param}
    \mathcal{Q}^{\mu} 
    \equiv&
    \frac{\tau_{Q,1}}{c^2} \left[
        a^{\mu} + \frac{c^2}{\rho h} \Delta^{\mu \nu} \nabla_{\nu} p 
    \right]
    \nonumber \\
    &+
    \frac{ \rho \kappa T^2}{m_b (e + p) c^2} 
    \Delta^{\mu\alpha} \nabla_{\alpha} \left( \frac{\mu}{T}\right)\, , \\
\label{eq:T-mu-nu-def-general-param}
    \mathcal{T}_{\mu \nu} 
    \equiv& 
    -2\eta\, \sigma_{\mu \nu}\,, \\
\end{align}
\end{subequations}
where $m_b$ denotes the baryon mass,  
$\zeta$ the bulk viscosity, $\eta$ the shear (dynamic) viscosity, 
and $\kappa$ the thermal conductivity.
The expansion $\theta$ and the shear tensor $\sigma_{\mu \nu}$ are defined by
\begin{align}
\label{eq:def_fluid_expansion}
\theta 
    &\equiv
    \nabla_{\mu}u^{\mu} 
    ,\\
\label{eq:def_fluid_shear}
\sigma^{\mu\nu} 
    &\equiv 
    \Delta^{\mu \gamma} \Delta^{\nu \delta} \nabla_{[\gamma} u_{\delta]} 
    -
    \frac{1}{3} \Delta^{\mu \nu}\Delta^{\gamma \delta} \nabla_{\gamma} u_{\delta}
    \,.
\end{align}

The functions $\tau_{\epsilon,i}, \tau_{p,i}, \tau_{Q,1}$ 
are transport coefficients which describe out-of-equilibrium contribution to the 
generalized energy, the generalized pressure and the heat flux vector.
They were introduced in \cite{Kovtun_2019,Bemfica:2020zjp}, and provided
they satisfy a set of inequalities, the equations of motion 
$\nabla_{\mu}T^{\mu\nu}=0,\nabla_{\mu}J^{\mu}=0$ 
form a strongly hyperbolic system of equations.
To obtain the BDN parameterization \cite{Bemfica:2020zjp}, we set
\begin{align}
   \tau_{\epsilon,1} &= \rho \tau_{\epsilon}\,,  \\
    \tau_{\epsilon,2} &= p \tau_{\epsilon}\,, \\
    \tau_{p,1} &= \tau_{p} \,, \\ 
    \tau_{p,2} &= p \tau_{p}\,,\\
    \tau_{Q,1} &= \tau_Q \rho h\,,
\end{align} 
and the rest of the coefficients to zero. One obtains Eckart's fluid model
if all of the $\tau_{\epsilon,i}$, $\tau_{p,i}$, and $\tau_{Q,1}$ coefficients
are set to zero.
\subsection{Israel-Stewart theory with bulk viscosity\label{sec:MIS-intro}}
In the (M\"{u}ller-)Israel-Stewart approach to relativistic, viscous fluid dynamics, the
higher gradient corrections in the fluid stress-energy tensor are promoted to their own 
independent variables, which satisfy new equations of motion 
\cite{Israel:1976tn,doi:10.1098/rspa.1977.0155,Israel:1979wp}\footnote{Sometimes
the Israel-Stewart class of models are known as \emph{second-order models}, to distinguish them
from \emph{first-order models} such as the BDNK model, where no auxiliary fields are added
\cite{Bemfica:2020zjp}.}.
There are a wide variety of viscous fluid models that apply the original ideas developed by
Israel and Stewart, including the DNMR \cite{Denicol_2012} and (r)BRSSS models \cite{Baier:2007ix}
fluid models, which have been extensively
applied in modeling the aftermath of heavy-ion collisions;
for a recent review see \cite{Romatschke:2017ejr}.
Here we focus on the only Israel-Stewart model that has been proven to be strongly hyperbolic
and causal for non-equilibrium, dynamical solutions \cite{Bemfica:2019cop}.
Moreover, this model was used recently in \cite{Most:2021zvc} to estimate the PN order at which 
viscosity would enter the waveform.

This particular variant of Israel-Stewart has the following stress-energy tensor 
and fluid current vector \cite{Bemfica:2019cop}
\begin{subequations}
\begin{align}
    T_{\mu\nu}
    &=
    \frac{e}{c^2}u_{\mu}u_{\nu}
    +
    \left(p+\Pi\right)\Delta_{\mu\nu}
    \\
    J^{\mu}
    &=
    \rho u^{\mu}
    .
\end{align}
\end{subequations}
Here $\Pi$ is a new, auxiliary field. The equations of motion are standard,
$\nabla_{\mu}T^{\mu\nu}=0$ and $\nabla_{\mu}J^{\mu}$, except for the addition of a new equation
of motion for $\Pi$ 
\begin{align}
    \label{eq:relaxation_is_bulk}
    \tau_{\Pi}u^{\alpha}\nabla_{\alpha}\Pi 
    +
    \Pi 
    +
    \lambda \Pi^2 
    +
    \zeta\theta 
    =
    0
    ,
\end{align}
where $\tau_{\Pi}>0$ is new relaxation time-scale, and $\lambda\geq0$  
is a new dimensionful constant.
From the form of \eqref{eq:relaxation_is_bulk}, we see that $\Pi$ should be driven towards 
$-\zeta\theta$ for near-equilibrium solutions.

\section{Stationary polar perturbation of the Einstein 
tensor\label{sec:stationary-polar-perturbation}}

Here we explicitly show that the heat vector $\mathcal{Q}_{\mu}$ does not enter
in the calculation of the $\ell=2$ polar tidal Love number, despite the fact that
in principle the $\mathcal{Q}_r$ component could be nonzero.
We note that the form of the Einstein equations remain unchanged for $\ell>2$, 
so our argument holds for $\ell>2$ as well.
The Einstein tensor components are
\begin{widetext}
\begin{align}
    \label{eq:pert_tt_comp_polar}
    \frac{\delta G^t_t}{Y^0_{2}}
    =&
    \frac{2}{r^2}K
    +
    \frac{e^{-2 \lambda } }{r}H_2' 
    +
    \frac{e^{-2 \lambda } \left(-3+r \lambda '\right)}{r}K' 
    +
    \frac{\left(3+e^{-2 \lambda }-2 e^{-2 \lambda } r \lambda '\right)}{r^2}H_2
    -
    e^{-2 \lambda } K''
    ,\\
    \label{eq:pert_rr_comp_polar}
    \frac{\delta G^r_r}{Y^0_{2}}
    =&
    -
    \frac{3 }{r^2}H_0 
    +
    \frac{2 }{r^2}K
    +
    \frac{e^{-2 \lambda } }{r} H_0' 
    +
    \frac{e^{-2 \lambda } \left(1-2 r \psi'\right)}{r^2}H_2 
    +
    \frac{e^{-2 \lambda } \left(-1+r \psi '\right)}{r}K' 
    ,\\
    \label{eq:pert_AA_comp_polar_tensor}
    \frac{
        \delta G^{\vartheta}_{\vartheta}
        -
        \delta G^{\varphi}_{\varphi}
    }
    {
        2 \left(
            \left[Y^0_{2}\right]^{\vartheta}_{\vartheta} 
            -
            \left[Y^0_{2}\right]^{\varphi}_{\varphi} 
        \right) 
    }
    =&
    \frac{1}{4 r^2}H_2 
    -
    \frac{1}{4 r^2}H_0 
    ,\\
    \label{eq:pert_AA_comp_polar_scalar}
    \frac{
        \delta G^{\vartheta}_{\vartheta}
        +
        \delta G^{\varphi}_{\varphi}
    }{
        2 \left(
            \left[Y^0_{2}\right]^{\vartheta}_{\vartheta} 
            +
            \left[Y^0_{2}\right]^{\varphi}_{\varphi} 
        \right) 
    }
    =&
    \frac{e^{-2 \lambda } \left(r \left(\lambda '+\psi '\right)-2\right)}{2 r}K'  
    -
    \frac{1}{2} e^{-2 \lambda }K''
    -
    \frac{3 }{2 r^2}H_0
    -
    \frac{
        e^{-2 \lambda } \left(r \lambda '+2 r \psi'-1\right)
    }{2 r}
    H_0' 
    +
    \frac{1}{2} e^{-2 \lambda } H_0''
    \nonumber\\
    &
    +
    \frac{
        \left(
            2 r e^{-2 \lambda } \left(
                \left(r \psi '-1\right) 
                \left(\lambda'+\psi '\right)-r \psi ''
            \right)
            +
            3
        \right)
    }{2 r^2}
    H_2  
    -
    \frac{e^{-2 \lambda } \left(r \psi '-1\right)}{2 r}H_2'  
    ,\\
    \label{eq:pert_rA_comp_polar_scalar}
    \frac{\delta G^r_{\vartheta}}{\left[Y^0_{2}\right]_{\vartheta}}
    =&
    -
    \frac{1}{2} e^{-2 \lambda } H_0'
    +
    \frac{1}{2} e^{-2 \lambda } K'
    +
    \frac{e^{-2 \lambda } \left(-1+r \psi '\right)}{2r}H_2 
    +
    \frac{e^{-2 \lambda } \left(1+r \psi '\right)}{2 r}H_0  
    .
\end{align}
\end{widetext}
In these expressions, we have factored out the dependence on the 
scalar ($Y^m_{2}$), polar vector ($\left[Y^m_{2}\right]_A$), 
and polar tensor ($\left[Y^m_{2}\right]^A_B$) spherical harmonics,
where $A$ indexes $\vartheta,\varphi$. Our notation for the spherical
harmonics follows \cite{Martel:2005ir}.
A prime (${}^{\prime}$) denotes a derivative with respect to $r$.

These are precisely the quantities that entered in the derivation of the
master equation for $H_0$ derived in \cite{Hinderer:2007mb}, which governs
the polar tidal response of the star.
We next outline the derivation of the master equation.
First, using $\delta T^{\vartheta}_{\vartheta} = \delta T^{\varphi}_{\varphi}$,
from Eq.~\eqref{eq:pert_AA_comp_polar_tensor} we conclude that $H_2=H_0\equiv H$. 
As $\delta T^r_{\vartheta}=0$, Eq.~\eqref{eq:pert_rA_comp_polar_scalar}
relates $H$ and $H^{\prime}$ with $K^{\prime}$. 
One can next use $\delta G^{\vartheta}_{\vartheta} + \delta G^{\varphi}_{\varphi}
=
\left(8\pi G/c^4\right)\left(\delta T^{\vartheta}_{\vartheta} + \delta T^{\varphi}_{\varphi}\right)
=
\left(8\pi G/c^4\right)\delta p$ to relate  $H$ to $\delta p$; see
Eq.~\eqref{eq:pert_AA_comp_polar_scalar}.
Finally, one can eliminate the $K$ in the $tt$ (Eq.~\eqref{eq:pert_tt_comp_polar})
and $rr$ (Eq.~\eqref{eq:pert_rr_comp_polar}) components of
the Einstein equations by subtracting them: 
$\delta G^t_t-\delta G^r_r= \left(8\pi G/c^4\right)\left(\delta T^t_t - \delta T^r_r\right)$.
One can eliminate $\delta e$ by using the equation of state, which 
(ignoring the baryon density $\rho$) is equal to $e\left(p\right)$, 
so $\delta e = \left(\partial e/\partial p\right)\delta p$.
The dependence on $\delta p$ can be removed Eq.~\eqref{eq:pert_AA_comp_polar_scalar}. 
At this point one is left
with an equation solely for $H$: the polar master equation, which we reproduce here
for the quadrupole ($\ell=2$) mode for completeness
\begin{align}
    \label{eq:master_eqn_h}
    &\frac{1}{r^2}e^{\psi+\lambda}\frac{d}{dr}\left(r^2e^{-\psi-\lambda}\frac{dH}{dr}\right)
    \nonumber\\
    &+
    \Bigg[
        -
        2\frac{d^2\psi}{dr^2}
        -
        2
        \left(\frac{d\psi}{dr}\right)^2
        +
        \left(-\frac{7}{r}+2\frac{d\lambda}{dr}\right)\frac{d\psi}{dr}
        \nonumber\\&\qquad 
        +
        \frac{3}{r}\frac{d\lambda}{dr}
        +
        \frac{1}{r}\left(\frac{d\psi}{dr}-\frac{d\lambda}{dr}\right)\frac{\partial e}{\partial p}
        -
        \frac{6}{r^2}e^{2\lambda}
    \Bigg]
    H
    =
    0
    .
\end{align}
We emphasize that this derivation made use of the 
$tt$, $rr$, $\vartheta\vartheta$, $\varphi\varphi$,
and $r\vartheta$ components of the perturbed equations of motion, 
and did not rely on $H_1$ being zero.
As the tidal response can be extracted from $\delta g_{tt}$
near spatial infinity \cite{Thorne:1997kt,Flanagan:2007ix,Hinderer:2007mb}, \eqref{eq:master_eqn_h} 
completely determines the adiabatic polar tidal Love number.

While $H_1$ does not enter in the calculation of the Love number,
we show it is zero outside of the star.
The nonzero tensor components that contain $H_1$ are
\begin{align}
    \frac{\delta G^r_t}{Y^0_{2}}
    =&
    -
    3\frac{e^{-2\lambda}}{r^2}H_1
    ,\\
    \frac{\delta G^{\vartheta}_t}{\left[Y^0_{2}\right]_{\vartheta}}
    =&
    -
    \frac{e^{\psi-\lambda}}{2r^2}\left(e^{-\psi-\lambda}H_1\right)^{\prime} 
    .
\end{align}
To complete our argument, we show that even if $H_1$ is nonzero inside the star,
it must be zero outside of the star, and thus it does not affect the asymptotic
metric. Outside of the star, $T^{\mu}{}_{\nu}=0$, so $G^{\mu}{}_{\nu}=0$.
This implies $\delta G^r{}{}_t=0$ outside of the star, and as $G^r{}_t\propto H_1$,
we see that $H_1=0$ in the exterior of the star.